\documentclass[fleqn,usenatbib]{rasti}

\usepackage[T1]{fontenc}

\DeclareRobustCommand{\VAN}[3]{#2}
\let\VANthebibliography\thebibliography
\def\thebibliography{\DeclareRobustCommand{\VAN}[3]{##3}\VANthebibliography}

\usepackage{caption}
\newcounter{definition}[section]

\usepackage[fleqn]{nccmath}
\usepackage{amsmath}	
\usepackage{amssymb}
\usepackage{awesomebox}
\usepackage{commath}
\usepackage{dirtytalk}
\usepackage{dsfont}
\usepackage{enumitem}
\usepackage{fixltx2e,fix-cm}
\usepackage{graphicx}	
\usepackage{imakeidx}
\usepackage{mathtools}
\usepackage{multicol}
\usepackage{siunitx}
\usepackage{soul}
\usepackage{subcaption}
\usepackage{tikz-dependency}
\usepackage{tikz}
\usepackage{xcolor}
\usepackage{listings}
\usepackage[automake, acronym, nohypertypes=acronym]{glossaries-extra}
\setabbreviationstyle[acronym]{long-short}

\newacronym{lsst}{LSST}{Legacy Survey of Space and Time at the Vera C. Rubin Observatory}
\newacronym{plasticc}{PLAsTiCC}{The Photometric LSST Astronomical Time-series Classification Challenge}
\newacronym{lstm}{LSTM}{long-short-term-memory}
\newacronym{cmb}{CMB}{Comsic Microwave Background}
\newacronym{nlp}{NLP}{natural language processing}
\newacronym{grus}{GRUs}{gated reticular units}
\newacronym{rnns}{RNNs}{recurrent neural networks}
\newacronym{pca}{PCA}{principle component analysis}
\newacronym{rnn}{RNN}{recurrent neural network}
\newacronym{cnn}{CNN}{convolutional neural network}
\newacronym{cnns}{CNNs}{convolutional neural networks}
\newacronym{csr}{CSR}{compressed sparse row}
\newacronym{csc}{CSC}{Compressed Sparse Column}
\newacronym{mnras}{MNRAS}{Monthly Notices of the Royal Astronomical Society}
\newacronym{rasti}{RASTI}{Royal Astronomical Society Techniques and Instruments}
\newacronym{mts}{MTS}{multivariate time-series benchmark dataset}
\newacronym{sne}{SNe}{Supernovae}
\newacronym{sn1}{SNIa}{Supernovae Type-Ia}
\newacronym{tsc}{TSC}{time-series classification}
\newacronym{wisdm}{WISDM}{Wireless Sensor Data Mining}

\makeglossaries

\usepackage{newtxtext,newtxmath}
\usepackage{breakurl}

\newcommand{\eg}{\emph{e.g.}}

\newcommand{\etc}{\emph{etc.}}
\newcommand{\hide}[1]{}

\newcommand{\idx}[1]{\emph{#1\index{#1}}}
\newcommand{\ie}{\emph{i.e.}}

\newcommand{\sw}[1]{{\texttt{#1}}}

\title[The Tiny Time-series Transformer]{The Tiny Time-series Transformer: Low-latency
High-throughput Classification of Astronomical Transients using Deep Model Compression}

\author[Allam Jr. et al.]{
Tarek Allam Jr.,$^{1}$\thanks{E-mail: tarek.allam.10@ucl.ac.uk}
Julien Peloton,$^{2}$
Jason D. McEwen$^{1, 3}$
\\
$^{1}$Mullard Space Science Laboratory, University College London, Holmbury St Mary, Dorking, Surrey RH5 6NT, UK\\
$^{2}$Université Paris-Saclay, CNRS/IN2P3, IJCLab, 91405 Orsay, France\\
$^{3}$The Alan Turing Institute, British Library, 96 Euston Rd, London NW1 2DB\\
}

\date{Accepted XXX. Received YYY; in original form ZZZ}

\pubyear{2023}

\begin{document}
\label{firstpage}
\pagerange{\pageref{firstpage}--\pageref{lastpage}}
\maketitle

\begin{abstract}

  A new golden age in astronomy is upon us, dominated by data. Large
  astronomical surveys are broadcasting unprecedented rates of information,
  demanding machine learning as a critical component in modern scientific
  pipelines to handle the deluge of data. The upcoming Legacy Survey of Space
  and Time (LSST) of the Vera C. Rubin Observatory will raise the big-data bar
  for time- domain astronomy, with an expected 10 million alerts per-night, and
  generating many petabytes of data over the lifetime of the survey. Fast and
  efficient classification algorithms that can operate in real-time, yet
  robustly and accurately, are needed for time-critical events where additional
  resources can be sought for follow-up analyses. In order to handle such data,
  state-of-the-art deep learning architectures coupled with tools that leverage
  modern hardware accelerators are essential.
  We showcase how the use of modern deep compression methods can achieve a
  $18\times$ reduction in model size, whilst preserving classification
  performance. We also show that in addition to the deep compression techniques,
  careful choice of file formats can improve inference latency, and thereby
  throughput of alerts, on the order of $8\times$ for local processing, and
  $5\times$ in a live production setting. To test this in a live setting, we
  deploy this optimised version of the original time-series transformer,
  \sw{t2}, into the community alert broking system of FINK on real Zwicky
  Transient Facility (ZTF) alert data, and compare throughput performance with
  other science modules that exist in FINK.
  The results shown herein emphasise the time-series transformer's suitability
  for real-time classification at LSST scale, and beyond, and introduce deep
  model compression as a fundamental tool for improving deploy-ability and
  scalable inference of deep learning models for transient classification.

\end{abstract}


\section{Inference in the Age of Large Synoptic Surveys}\label{section:finkmod-intro}

The turn of the century has seen a move towards ever larger astronomical
surveys, collecting large volumes of synoptic data across the night-sky, as
opposed to previous instruments that focus data collection for a single science
case. Being able to conduct a large swath of science from a single data source
is one of the main drivers for development and construction of such surveys, and
allows for many science communities to benefit from a single instrument.
Recent surveys such the Sloan Digital Sky Survey~\citep[SDSS][]{york2000sloan},
the Dark Energy Survey~\citep[DES][]{abbott2016dark}, the Panoramic Survey
Telescope and Rapid Response System~\citep[Pan-STARRS][]{kaiser2002pan} to name
a few, are examples of astronomical surveys that map the sky without a
particular astronomical \emph{target} in mind. They are often limited in scope
in terms of electromagnetic spectrum, but can serve as the precursor to more
specialised instruments for follow-up observations. Typically, surveys are used
to generate catalogues of astronomical objects, as well as logging astronomical
transient events on the sky.

The detection of transient events is of particular importance for probing the
nature of dark energy and constraining theories of the
Universe~\citep{riess1998observational, perlmutter1999measurements}.
Typically SNe, which are observed over a period of a few days to a few weeks,
are classified by the presence of particular absorption lines in their spectra.
Specifically, \gls{sn1} are distinguished by the absence of hydrogen lines and
the presence of Si II $\lambda$ 6150 absorption~\citep{riess2001farthest}.
However, spectroscopic classification of transient events is a costly process.
By using broad photometric passbands, LSST will be able to \say{see} far more
events than ever before, or that could be possible with spectroscopic equipment.
The problem then arises: how can one accurately identify different transient
photometrically using only passband information?
In contrast to spectroscopic classification, photometric classification is far
more challenging, and one is more susceptible to cross-contamination from other
events such as core-collapse SNe (SNe Ib/c and SNe II) which share a similar
profile to SNIa when observed photometrically.
Consequently, studies have been done by similar photometric surveys to determine
the acceptable level of cross-contamination from such events that would still
allow for robust cosmological analysis of the dark energy equation of state.
This range has been reported to be between $8\%$~\citep[DES][]{vincenzi2021dark}
and $5\%$~\citep[Pan-STARRS][]{kaiser2002pan}.
It is expected LSST will require a high SNIa purity and cross-contamination rate
to be at least within this range, if not lower. It should be noted that these
levels are in the context of full phase light curves, and so one may expect a
higher level of cross-contamination in the early phase of the events, where only
partial information is available for identification.

When making observations photometrically, the flux measurement corresponding to
a given passband is obtained by collecting all light that is received though
that particular filter. Multi-band photometry allows for more information to be
retrieved to help determine properties of the light source, such as temperature,
but this is of course not as rich as observing spectroscopically. However, if
one collects multi-band photometry over a period of time, of the same source, a
light curve can be constructed, which tells us more about what kind of a
transitive event this may be.

The image processing method of difference imaging measures differential
photometry by matching the pointing and point-spread function(s) between image
frames, typically for the detection time-varying celestial
objects~\citep{wang2017pixel}.
A new image is compared with an aligned reference image, where the difference
between the two images is determined by calculating the difference between each
pixel of each image, and forming a difference image from the result. A detection
occurs when the difference image is above a certain signal-to-noise threshold.
When this threshold is reached, a transient event alert is triggered, with data
streamed to brokers around the world for follow-up analysis.

With difference imaging processing now done entirely in software, the need to
automate pipelines for detection and classification of transient events comes
from the sheer volume of data these surveys generate, as well as the number of
events they witness. Developments in instrumentation have allowed these surveys
to scan larger areas of the sky and more detailed than ever before, with the
consequence being that machine learning has now become a critical component in
order to process the data.

When the Legacy Survey of Space and Time (LSST) at the Vera C. Rubin Observatory
comes online, it is expected to observe 10 million variable and transient events
per night, generating on the order of 1TB of data per night\footnotemark. The
tsunami of transient alerts that is to be distributed globally, calls for
machine learning systems that can scale to such data rates, and yet still
provide robust identification of events. A classifier that can process an alert
and provide classification scores in real-time will not only enable efficient
allocation of resources for follow-up observations, but assist with labelling of
the millions of events which is certainly beyond humans at this point.
At this scale, storage space and computational costs becomes a real concern, and
so for real-time processing and machine learning enriched catalogues to be
feasible, classification modules need to be lightweight in terms of storage
space and runtime memory footprint. Furthermore, computations should be done as
efficiently as possible to not only save on time, but also money by minimising
energy usage.

\footnotetext{Raw data volume for image and calibration data will not be
distributed with the alerts to save space and reduce the data sent over the
wire, which would otherwise amount to 60PB over the course of 10 years of
operation.}

We structure this article in the following way.
Section~\ref{section:broker-call} discusses the challenges involved when dealing
with a large influx of data from space surveys, and stresses the need for alert
brokering systems. We touch on the policies in place for developing such brokers
and describe what motivates the use of the Zwicky Transient
Facility~\citep[ZTF][]{bellm2014zwicky} alert stream in preparation for the
upcoming LSST data distribution in Section~\ref{section:proxy}.
Following that, our focus turns to one broking system in particular,
FINK~\citep{moller2021fink}, which served as the platform for the research
discussed later in this article. Section~\ref{section:performance-engineering}
explains the ideas behind the engineering efforts that ultimately allowed for
our deep learning model to successfully operate in a production system. Our
preliminary results on local tests are showcased in
Section~\ref{deploy:section:preliminary-results}, followed by the results of
applying our methods in real-world conditions in
Section~\ref{deploy:section:production-results}. We then conclude in
Section~\ref{deploy:section:conclusions} with a discussion of these results, and
how use of the methods described facilitate efficient deep learning and
real-time inference, at LSST scale.

\section{Calling All Brokers!}\label{section:broker-call}

Due to computational constraints, and limits on bandwidth, the full distribution
of alerts from LSST Data Facility\footnotemark~will only be sent out to a select
few community brokering systems, whose primary purpose is to provide catalogue
cross-match functionality and photometric classification of objects, thereby
enriching the raw alert packets with value-added information for downstream
scientific investigations.
\footnotetext{\href{https://www.lsst.org/about/dm/facilities}{lsst.org/about/dm/facilities}}
A worldwide call for brokering systems was announced in 2019 to entice teams
interested in potentially building such systems~\citep{LDM-682}, which was soon
followed by a call for concrete proposals the following year~\citep{LDM-723}.

\subsection{Community Alert Brokers}\label{section:comm-brokers}

Since the full alert stream can not be accessed directly by scientists,
community brokering systems are essential software systems that will enable
time-domain science~\citep{LDM-612}. Further to the requirements of
cross-matching and photometric classification, brokers are also expected to
provide additional services to enable science such as a simplified user
interface for easy querying of archival data, a triggering follow-up observing
service, additional alert filtering\footnotemark, among others.
The call for brokers was not limited to any institutions in particular, and the
open call encouraged a variety of system designs. As long as there is capacity
for petabytes (PB) of storage space, a large inbound bandwidth network,
real-time machine learning classification modules, and of course sufficient
funding, brokering teams were free to set out their plans in The Call for
Proposals for Community Alert Brokers~\citep{LDM-723}.
Naturally, brokers that offer a suite of services along with the necessary
infrastructure capabilities, were seen more favourably by the selection
committee, and in particular brokers that take advantage of the unique real-time
aspects of the LSST alert stream~\citep{LDM-723}. Moreover, brokers that already
exhibit integrations with follow-up resources and other surveys through existing
agreements, as well as scope for community building, was also viewed positively.

\footnotetext{The LSST Data Facility applies its own filtering of alert packets
such as a criterion of SNR $>5$ and before distribution}

Of the many teams that put forward proposals, seven teams were ultimately chosen
that showcased their ability to match the criteria laid out above, or at least
showed the potential to realise the requirements come time of first light. The
successful broker bids were from teams; The Automatic Learning for the Rapid
Classification of Events (ALerCE)~\citep{forster2021automatic},
AMPEL~\citep{nordin2019transient}, Arizona-NOAO Temporal Analysis and Response
to Events System (ANTARES)~\citep{matheson2021antares},
BABAMUL~\citep{babamul-broker}, Pitt-Google~\citep{pitt-google-broker},
Lasair~\citep{smith2019lasair} and FINK~\citep{moller2021fink}.

\subsection{The ZTF Alert Stream: A Proxy for Success}\label{section:proxy}

In order to support development of the broking systems, LSST provided example
alert streams to get broking teams familiar with the expected data formats and
protocols. Much of these were inspired by the Zwicky Transient Facility (ZTF)
already-in-action alert distribution system~\citep{DMTN-093}.
The Zwicky Transient Facility (ZTF) is an astronomical survey that observes in
visible and infrared, primarily focusing on the detection of transient objects
that change rapidly~\citep{bellm2014zwicky}. Its high cadence allows it to
observe the entire northern sky in three nights over two passbands\footnote{Only
\emph{g} and \emph{r} filter bands are available from the public alert stream,
while the \emph{i} filter band is only accessible to the private part of the
survey.}. Although generating only a tenth of the data expected from the LSST,
the data pipelines and alert distribution systems in place with ZTF have been
shown to be suitable to act as a precursor to the much larger data rates of
fully operational LSST~\citep{patterson2018zwicky}.

The ZTF alert packets are in \sw{Apache Avro}\footnotemark~format, a binary
serialisation format, that contains difference imaging information of the
detection, yet is still compact and lightweight enough for real-time worldwide
publishing.
\footnotetext{\href{https://github.com/apache/avro}{github.com/apache/avro}}
Deserialisation is done in conjunction with a corresponding alert schema that
defines the contents of the alert packet, and hence the information that can be
used for processing. LSST is set to follow the same data format as well as the
same \emph{pub-sub} framework using \sw{Apache Kafka}\footnotemark~for the
distribution of the streaming data, where it collects data streams at source,
from \emph{producers} (\eg~the telescope itself) and arranges them into sets of
\emph{topics} that can be subscribed to by \emph{consumers} downstream
(\eg~community alerts brokers).
\footnotetext{\href{https://github.com/apache/kafka}{github.com/apache/kafka}}

Since 2018, the ZTF alert production system produces on average 250 thousand new
alerts nightly, and it has shown to successfully support streams of 1.2 million
nightly alerts, which equates to approximately 70 GB per night, where each alert
packet has been made available within 10 seconds of event
detection~\citep{patterson2018zwicky}.
On the order of 80,000 alerts per minute, the stability of the production system
when dealing with such data rates gives support to the case for using the same
technologies and protocols described in~\citet{patterson2018zwicky} for
developing brokers that are to be suitable for the $10\times$ larger upcoming
Large Survey of Space and Time (larger in terms of number of alerts, but also in
terms of alert size).

\subsection{FINK:\ A Next Generation Broker}\label{section:fink}

Of the seven brokers mentioned in Section~\ref{section:comm-brokers} that were
successful in securing a spot as one of the brokering systems, we discuss FINK
in more detail here. FINK is the system that the modified deep learning model
of~\citet{allam2021paying} was deployed to, and was the ultimate test bed for
investigating the real-time capabilities of the original time-series
transformer~\citep{allam2021paying}.
FINK goes beyond typical brokering systems by providing real-time transient
classification through fast state-of-the-art deep learning algorithms that can
be re-trained at low cost in a short space of time, and by using active learning
techniques, specifically online learning, that allow for continuous
self-improvement of classification scores.
Specifically designed to address the challenges outlined in
Section~\ref{section:finkmod-intro}, it uses industry standard tools for
efficient big data processing. In order to carry out nightly processing of the
terabyte data stream, FINK uses fault-tolerant \sw{Apache
Spark}~\citep{zaharia2016apache} for scaling out computations across many
computers, and Spark Structured Streaming~\citep{armbrust2018structured} to
easily interact with \sw{Apache Kafka} for \emph{consuming} the data stream.

\begin{figure*} \centering
  \includegraphics[width=0.9\textwidth]{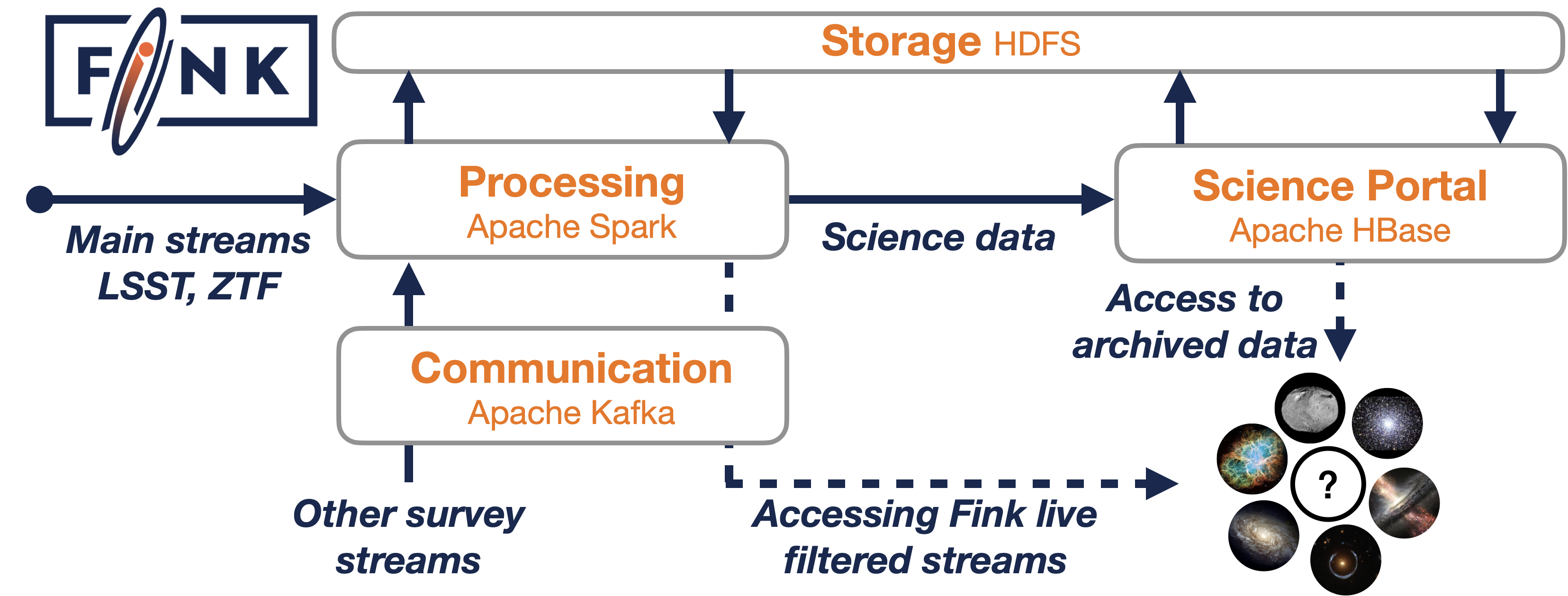}
  \caption[FINK pipeline and system architecture.]{FINK pipeline and system
  architecture, where the main alert streams are processed in a distributed
  manner using \sw{Apache Spark}~\citep{zaharia2016apache} on a Processing
  cluster. Following the initial processing, a set of sub-streams are produced
  which users can subscribe to by way of the Communication cluster using
  \sw{Apache Kafka}. Further survey data streams, such as those from LIGO,
  Fermi, \etc~are ingested into the Processing cluster through the Communication
  cluster, to enrich alert packets with added information. This is used in
  conjunction with science modules within the Processing cluster that provide
  classification scores for alerts and added-value information. After a night of
  operation, the processed and enhanced data is written to the Hadoop
  Distributed File System (\sw{HDFS}) in the Storage cluster which connects to a
  Science Portal backed by the distributed database of \sw{Apache
HBase}\footnotemark~to allow for interactive querying of archived alerts.
Reproduced in full from~\citet{moller2021fink}} \label{figure:fink-infra}
\end{figure*}

FINK currently has a memorandum-of-understanding (MoU) with the Zwicky Transient
Facility (ZTF), allowing it to receive real alert packet data, in the form
described in Section~\ref{section:proxy}, each night. This makes FINK well
suited for not only testing how well deep learning models can perform under
stress with real-time constraints, but also to test how well models handle
classification of \emph{real} data.
The FINK system diagram can be seen in Figure~\ref{figure:fink-infra}. Along
with mapping the flow of data through the broker, Figure~\ref{figure:fink-infra}
also shows at which stage the redistribution of enriched alerts will take place.
The interplay of the photometric classification modules in the Processing
cluster, additional third-party survey data via the Communication cluster and
aggregation of value-added information in the Storage cluster, form the
foundations of the Science Portal, which can be used to enable scientific
analysis and offline querying of the archival data for those in the community.
As of 2023, the Science Portal gives access to more than 10 TB of alert data.

\footnotetext{\href{https://github.com/apache/hbase}{github.com/apache/hbase}}

\section{The Time-series Transformer}\label{section:t2}

This section reviews the time-series transformer architecture
of~\citep{allam2021paying} that motivated this work and highlights the appeal of
using such an architecture for fast and efficient astronomical transient
classification.

\subsection{The Motivational Driver}

The problem of time-series classification is one that extends to a vast number
of disciplines. As with many domains, traditional approaches involved
hand-crafted feature engineering to uncover patterns that would be useful for
classification. Today, with the shear volume of data, deep learning methods are
being investigated as a promising alternative to previous methods for
classification~\citep{fawaz2019deep}.

Of the deep learning methods applied to general time-series classification
discussed in~\citet{fawaz2019deep}, the most successful architectures have been
variants of convolutional neural networks (CNNs) with work
by~\citet{wang2016time} and by~\citet{geng2018cost}.  In the field of astronomy,
recurrent neural networks (RNNs) have become popular for astronomical transient
classification~\citep{moller2020supernnova, muthukrishna2019rapid,
charnock2017deep}.

Although these deep learning methods achieve impressive results, both RNNs and
CNNs often struggle when dealing with time-series data of increasing sequence
length.
The inherently sequential structure of RNNs makes parallelisable computation
troublesome since each input point needs to be processed in order.
CNNs, on the other hand parallelise easily and, with the use of the dilated
convolution, larger sequences can be processed \citep{oord2016wavenet}. Having
said that, CNNs suffer greatly from being relatively computationally expensive
compared to RNNs~\citep{vaswani2017attention}.
The transformer architecture and accompanying self-attention mechanisms help
mitigate these problems faced by CNNs and RNNs for long-sequence time-series
data~\citep{vaswani2017attention}. The work by~\citet{allam2021paying} proposed
a new transformer architecture for time-series classification that was
originally motivated by astronomical transient classification but was found to
be versatile enough to be applied to general multivariate time-series data.

\begin{figure} \centering \hspace*{-1.5cm}
  \makebox[0pt]{\includegraphics[height=13.25cm]{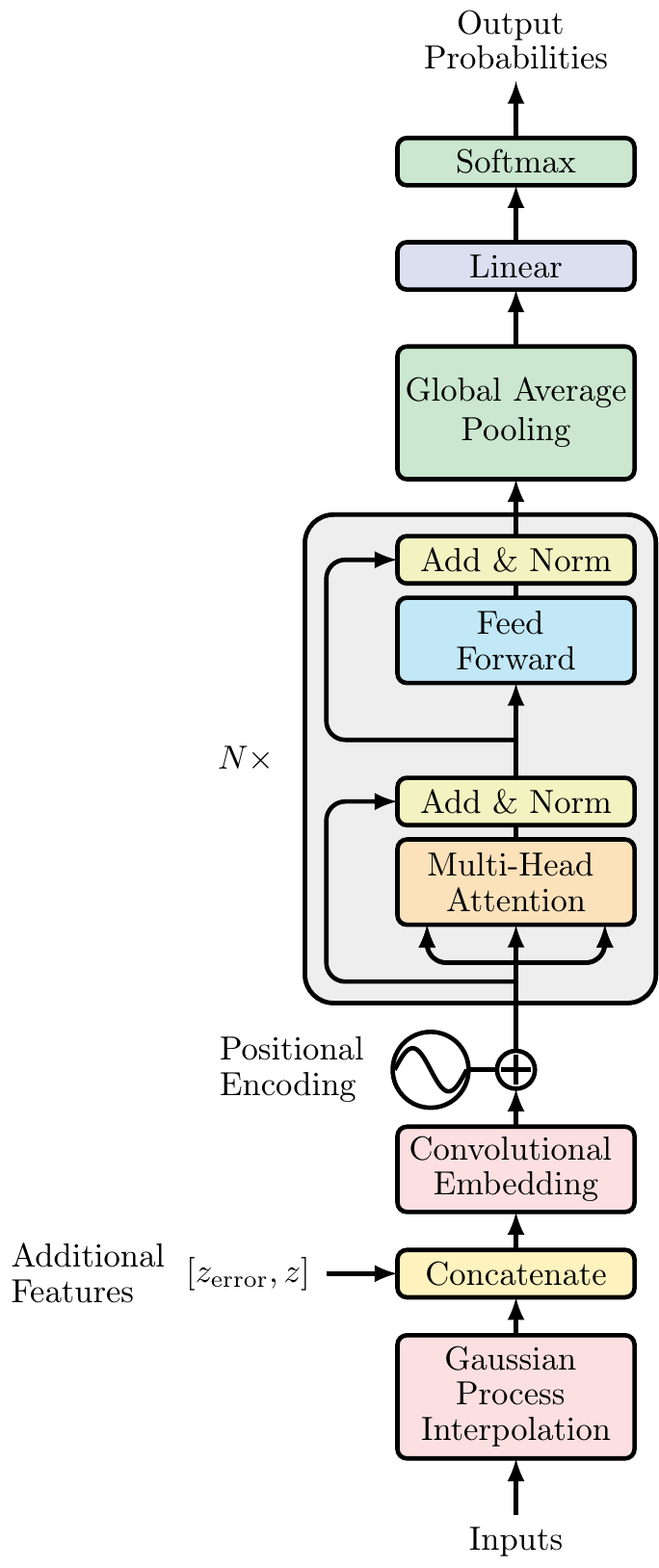}}%
  \caption{Schematic of the time-series transformer (\sw{t2})
  architecture~\citep{allam2021paying}. If the time-series signal is irregularly
  sampled, then data is passed through the Gaussian process interpolation layer,
  followed by a concatenation layer to include any additional features. The
  embedding layer follows to transform the input, with a positional encoding
  applied to this embedding vector. The following multi head attention block is
  the same as that shown in~\citet{vaswani2017attention}. A GAP layer is then
  applied and finally a linear layer with softmax to output class prediction
  probabilities for the objects. Reproduced in full
from~\citet{allam2021paying}.} \label{figure:t2} \end{figure}

As described in~\citet{allam2021paying}, the time-series transformer, \sw{t2},
shown in Figure~\ref{figure:t2}, forgoes the decoder arm of the traditional
transformer architecture of~\citet{vaswani2017attention} yet adds new layers for
Gaussian Process interpolation and inclusion of additional features.
The few number of parameters used in this model compared to other deep learning
approaches, combined with fast and parallelisable nature of the attention
mechanisms within make for an appealing architecture to be deployed. In the next
section we discuss the performance engineering choices that were made to improve
the time-series transformer's capabilities even further, placing it into the
realm of production ready real-time systems operating at scale.

\section{Performance Engineering for Deployment in FINK}\label{section:performance-engineering}

Since photometric classifiers are to be housed in the Processing cluster (see
Figure~\ref{figure:fink-infra}), low-latency, high-throughput algorithms are
essential to handle the deluge of data that is to be processed. This section
describes the research and development of a multistage compression process in
order to ensure best classification performance, while optimising for
low-latency inference and high-throughput processing of alerts.

\subsection{Line Profile Analysis}\label{deploy:section:line-profile-analysis}

In order to better understand the bottlenecks in our deep learning pipeline from
alert packet processing to inference, as well as to investigate the overall
systems performance, it is necessary to conduct a form of dynamic program
analysis. In contrast to static program analysis, which evaluates a program
without execution, dynamic program analysis focuses on the program's memory
usage, time complexity, duration of function calls \etc~
Such analyses are typically done through unit and integration tests, which
themselves may include, or can exist separate to the main codebase and line
profiling tests that scrutinise a program line-by-line. By observing time spent
at each function call, one can see where in the pipeline optimisations can be
made, and as such apply performance engineering techniques that reduce runtime
and memory footprint of the program.

To run such a test, we simulate locally the full pipeline from ingestion of a
real ZTF alert packet, to interpolated time-series, to model predictions, and
gauge where optimisations should be applied by using the line profiling tool
\sw{kernprof}\footnotemark. The line profiling software reports the time spent
on each function and the number of times that function has been called, for each
line of code in a Python program.

\footnotetext{\href{https://github.com/pyutils/line\_profiler}{github.com/pyutils/line\_profiler}}

Although initial expectations were that the main bottleneck would be the time-series interpolation
through Gaussian processes, the major bottleneck in the pipeline was found to be loading our model
into memory and applying the model to the input data for predictions (see
Listing~\ref{listing:initial-bottlenecks}). To combat this, we looked into ways of reducing model
size for faster model loading and operational changes to improve runtime latency.

\begin{lstlisting}[
  language=bash,
  caption={Line profile report for initial run of alert packet pipeline. Superfluous lines that
  recored less than $0.1\%$ time are removed for better readability. Note the function to generate
  the Gaussian process only takes $9.5\%$ of the total time, with the majority of time taken up with
  model predictions. For the full report, see
  {\href{https://github.com/tallamjr/astronet/blob/master/astronet/tests/reg/ztf-load-run-lpa.py.stdout.txt}{github.com/tallamjr/astronet/astronet/tests/reg}}},
  label=listing:initial-bottlenecks
]

Total time: 5.85664 s
File: get_models.py
Function: get_model at line 29

Total time: 1.47076 s
File: ztf-load-run-lpa.py
Function: t2_probs at line 55

Line #      Hits         Time  Per Hit   \% Time  Line Contents
==============================================================
...
206        16        139.6      8.7      9.5      df_gp_mean = generate_gp_all_objects()
...
...
...
...
...
212         8        180.8     22.6     12.3      X = df_gp_mean[cols]
213         8         12.3      1.5      0.8      X = rs(X)
...
...
...
217         8       1101.7    137.7     74.9      y_preds = model.predict(X)
\end{lstlisting}



\subsection{\emph{Deep} Compression}\label{section:deep-compression}

With the major bottleneck for fast inference identified to be at the model load
and then prediction stage, we focus our attention to model optimisations that
can be applied to alleviate this. Since our desire is to run the complex model
in real-time, we look to exploit redundancies in the model, thereby reducing the
storage size, lowering inference latency, and improving energy efficiency
processing alerts.
A relatively recent area of research that looks to reduce model size and memory
footprint of deep learning models it that of \idx{deep
compression}~\citep{han2015deep_compression}. Originally proposed as a
three-step process to reduce the computational cost and memory usage of deep
networks on embedded devices, it is mostly driven today by the interests of
industry for deploying deep learning models in-the-wild on resource constrained
devices. This influential work saw a new field flourish that combines bit saving
best-practises with deep learning architecture design to reduce storage size,
whilst at the same time preserving model accuracy.

For our research, we restrict our investigation to the techniques broadly laid
out in~\citet{han2015deep_compression}, namely \idx{weight-pruning} and
\idx{weight-clustering} with Huffman encoding, and \idx{weight-quantization}.
All of which can be applied in isolation or together, with the caveat being that
if these techniques are chained together, the possibility for severe degradation
in performance is high.

\subsubsection{\emph{Pruning}}

Pruning is a technique that removes unimportant weights to yield improvements
such as better generalisation and improved speed of learning and
classification~\citep{lecun1989optimal}. It has been shown in recent times that
deep neural network can be pruned to a significant degree with little reduction
in model accuracy~\citep{han2015learning, han2015deep_compression}.

While there are many forms of network pruning such as
layer-pruning~\citep{lazarevich2021post}, channel-pruning~\citep{he2017channel},
filter-pruning~\citep{enderich2021holistic} and
connection-pruning~\citep{nguyen2021connection}, we consider magnitude-based
weight pruning here~\citep{han2015deep_compression}, where the weights are
updated network-wide through a small number of fine-tuning epochs to zero out
model weights that have a low impact on the final score, creating a sparse
representation of the model.
Sparse models\footnotemark~can then leverage standard lossless compression
algorithms for large reduction in model size, as well as faster inference
through fewer parameters and hence fewer computations. \footnotetext{Stored in
\gls{csr} or \gls{csc} format gives $2a + n + 1$ numbers, where $a$ is the
number of non-zero elements, and $n$ is the number or rows or columns, which is
normally $ \ll L \times M$ size of a complete matrix of all elements}

\subsubsection{\emph{Clustering}}

Another method that promotes sparsity in the network is through
weight-clustering. Also commonly referred to as weight-sharing, clustering works
by grouping the weights of each layer in a model into a predefined number of
clusters. The centroid values for the clusters are then shared among the weights
in the given cluster. By dividing the $m$ original number of weights in the
network, $W = \{w_1, \dots, w_m\}$ into $k$ clusters $C = \{c_1, \dots, c_k \}$,
where $m \gg k$ there is a great reduction in the number of \emph{unique} weight
values in a model, which in turn allows for greater storage efficiency and high
data compression potential.
If we consider there to be $n$ possible connections in the network, where each
connection is represented by $b$ bits, then a fully connected network would be
represented with $n \cdot b$ bits. A clustered network on the other hand, with
$k$ clusters, requires only a cluster encoding index of $\log_{2}(k)$ together
with the number of clusters with shared weights. This yields a compression rate,
$r$, of

\begin{align} r=\frac{n \cdot b}{n \cdot \log_{2}(k)+k\cdot b}.
\label{equation:compression-rate} \end{align}

\begin{figure}
    \centering
    \includegraphics[width=0.45\textwidth]{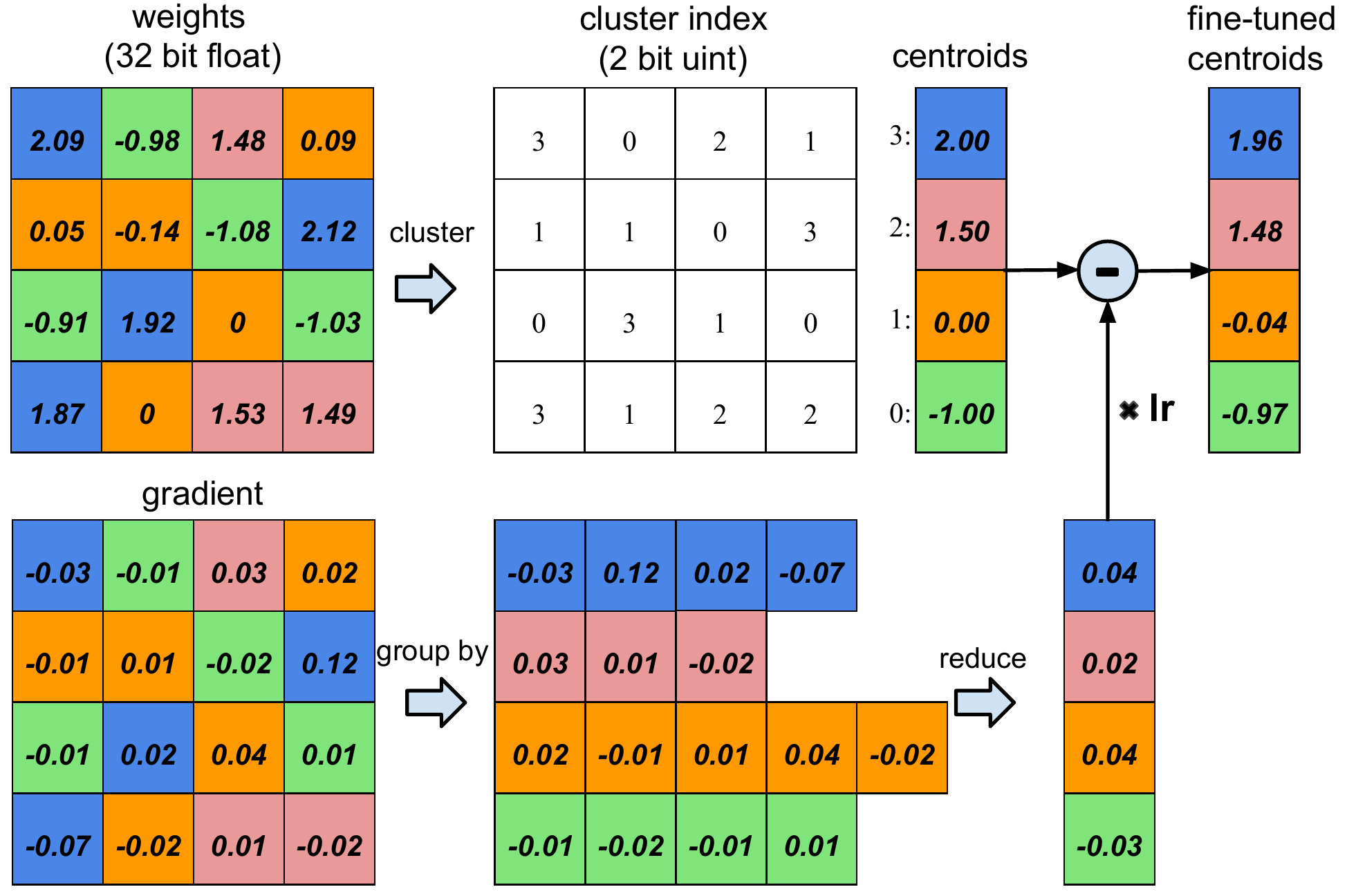}
    \caption[Weight clustering compression scheme.]{Weight clustering
    compression scheme, showing the weights of a single layer neural network
    that has four input and four output units. In total there are 16 weights,
    which are reduced to a set of 4 shared weights. The top row depicts the full
    weight matrix for the 4 by 4 input-output connections, whereas the bottom
    row shows the related gradient matrix. As an example, using 4 colours to
    denote the 4 clusters, the set of weights are put into one of 4 clusters,
    where all values in the same cluster share the same value. As such, an index
    mapping the weight to a particular cluster is stored. In the training phase,
    the gradients are grouped by colour (cluster), summed, multiplied by the
    learning rate, and finally subtracted from the shared centroids of the last
    iteration. Reproduced in full from~\citet{han2015deep_compression}}
    \label{figure:clustering}
\end{figure}

An example using a single fully connected four-by-four neural network can be see
in Figure~\ref{figure:clustering}. If we use
Equation~\ref{equation:compression-rate}, we can see that by setting $k = 4$
(using 4 distinct colours to signify separate clusters), one is able to reduce
required number of bits for the original 16 weights at 32 bit precision, down by
a notable factor with a compression rate of 3.2.

\begin{figure}
    \centering
    \hspace*{-0.5cm}
    \includegraphics[width=0.45\textwidth]{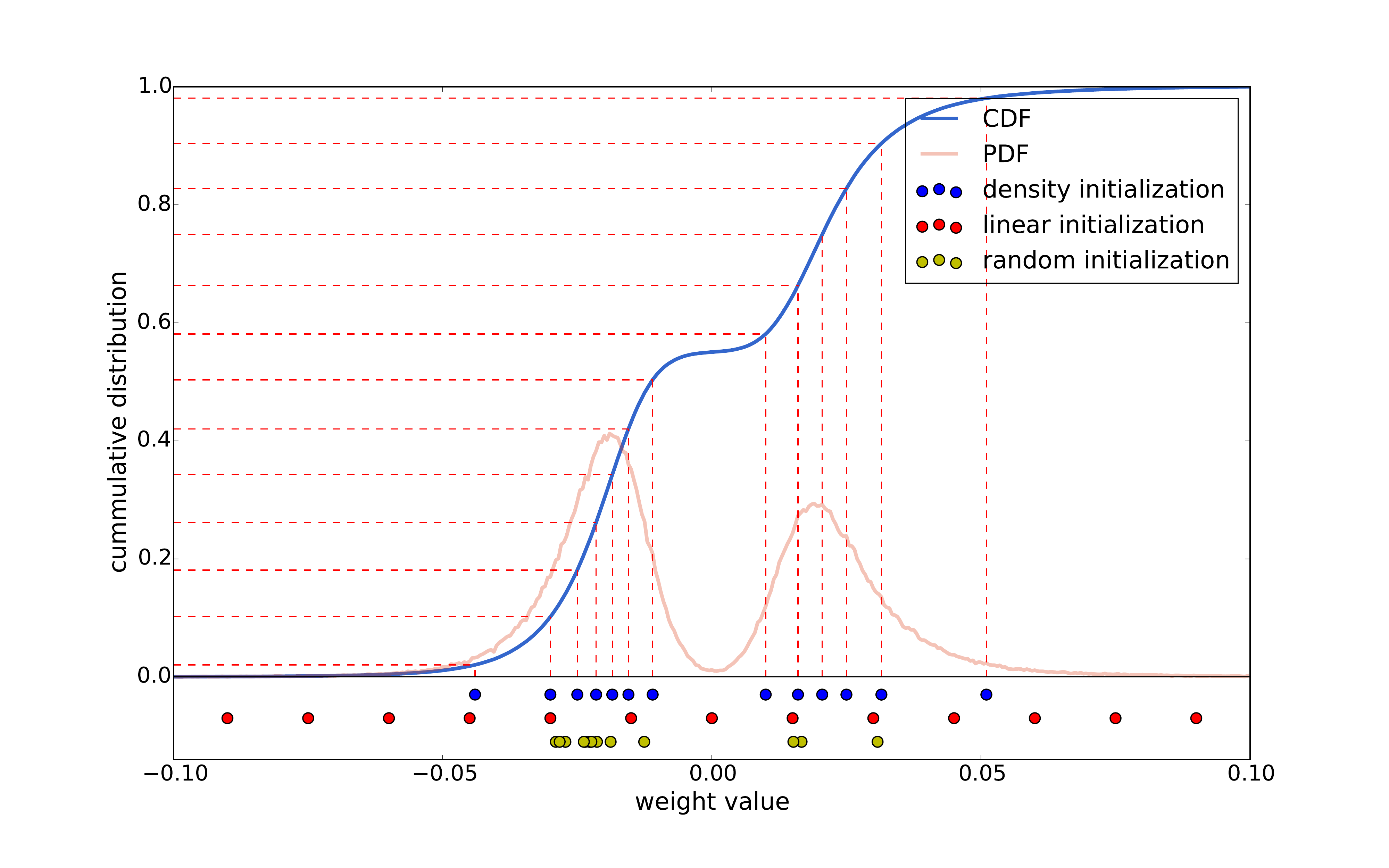}
    \caption[Centroid initialisation schemes.]{Centroid initialisation schemes,
    using the bimodal distribution of weights in CONV3 layer of
    AlexNet~\citep{krizhevsky2012imagenet} as an example. Shown at the bottom
    are the 13 cluster centroids for this example layer using 3 different types
    of centroid initialisation schemes. {\bf{Random}}: randomly selects $k$
    points from the data set and uses these as the initial centroids, shown in
    yellow. {\bf{Density}}: uses the cumulative distribution of the weights to
    create a linear spacing on the $y$-axis, and then finds the corresponding
    $x$-axis value that intersects with the distribution, shown in blue.
    {\bf{Linear}}: evenly spaces the $x$-axis of weights from min to max value
    and then places a centroid, shown in red. Random initialization tends to
    concentrate around the peaks of the distribution, as does density-based
    initialization, albeit more scattered. Linear is even more scattered, but to
    note the initialization scheme is invariant to the weight distribution.
    Reproduced in full from~\citet{han2015deep_compression}}
    \label{figure:centroid-init}
\end{figure}

The canonical $k$-means clustering algorithm is used to find the clusters, but
of great importance in terms of the eventual model accuracy is how the centroids
are initialised. Of the three methods for centroid initialisation
in~\citet{han2015deep_compression}, random, density-based and linear, the
authors report that random and density-based centroid initialisation achieve
poor performance due to few centroids having large absolute values. Linear
initialisation on the other hand does not suffer from this problem, and is shown
by~\citet{han2015deep_compression} to work best under various conditions.
A comparison of the different centroid initialisation schemes is shown in
Figure~\ref{figure:centroid-init}.
For training, a weight lookup table is necessary to maintain information about
the shared weights and their connections among the clusters. The gradient for
each shared weight is then calculated and used to update the actual shared
weight, as can be seen in Figure~\ref{figure:clustering}. The gradient of the
centroids is given by,

\begin{align}
  \frac{\partial \mathcal{L}}{\partial C_{k}}=\sum_{i, j} \frac{\partial \mathcal{L}}{\partial W_{i j}} \frac{\partial W_{i j}}{\partial C_{k}}=\sum_{i, j} \frac{\partial \mathcal{L}}{\partial W_{i j}} \mathds{1} \left(I_{i j}=k\right)
  \label{equation:shared-weight-grads}
\end{align}
\noindent where $\mathcal{L}$ is the loss, $C_k$ as the $k$-th centroid, and the centroid index of
the weight matrix $W_{ij}$ is $I_{i j}$.

At the stage for which the model is to be deployed, \ie~for inference operations
only, the intermediate weight table can be \emph{stripped} from the model to
leave just the clustered weights, suitable for standard data compression
algorithms to reduce model size on disk.

\subsubsection{\emph{Quantization}}


Quantization is the simple procedure of reducing the number of bits used for
representing numbers. Weight-quantization helps reduce the storage and
computational requirement of the model, and in the case of our discussion here,
applied after training is completed\footnotemark. Post-training quantization
refers to the application of quantization by statically mapping the weight
values to lower precision integers, where in this lower precision
representation, the weights save significant amount of space on disk, and can
even see improvements to latency by leveraging efficient integer kernel
operators found in modern hardware accelerators.

\footnotetext{Quantization aware training on the other hand is a quantization
procedure that is applied during training by way of integer arithmetic for
computations.}

\subsection{Lossless Data Compression}\label{deploy:section:lossless}

As touched on in our brief overview of deep compression, many of the techniques
lend themselves well to exploitation by standard lossless data compression
algorithms. Both pruning and clustering induce redundancies in the model through
repeated values. The canonical compression scheme to handle repeated values is
Huffman encoding~\citep{huffman1952method}, which assigns fewer bits to repeated
values.
As such, it is recommended to combine sparsity inducing methods of deep
compression with lossless data compression algorithms. We use the DEFLATE
algorithm~\citep{deutsch1996rfc1951} within \sw{zlib}\footnotemark~which
combines Huffman encoding with the LZ77 compression
algorithm~\citep{ziv1977universal} to realise the full benefits of sparsifiying
our network. Though, we are mindful of the potential trade-off between storage
space savings when using lossless compression tools and the inevitable higher
latency caused by the decompression overheads when loading models into memory.
\footnotetext{\href{https://github.com/madler/zlib}{github.com/madler/zlib}}

\subsection{Efficient File Formats and Frameworks}\label{section:file-formats}

While application of deep compression techniques are likely to significantly
reduce the size of our deep learning model on disk, the possible increase in
latency times in relation to decompression overheads spurred an off-shoot
evaluation which looked at alternative file formats and lighter frameworks that
could help improve runtime of model predictions.

The work of~\citet{allam2021paying} (\sw{t2}) use
\sw{ProtocolBuffers}\footnotemark~as the serialisation format for saving models
developed using the full TensorFlow framework~\citep{tensorflow2015-whitepaper}.
\footnotetext{\href{https://github.com/protocolbuffers/protobuf}{github.com/protocolbuffers/protobuf}}
Inspired by the \index{TinyML} movement~\citep{david2021tensorflow}, that seeks
to run deep learning on \emph{extremely} resource constrained devices, we look
at the possibility of using only a subset of the full TensorFlow framework,
called TensorFlow Lite (TFLite)~\citep{li2020tensorflow}. Compared to the some
1400 operations in the full framework, TFLite only has around 130 operations
supported~\citep{david2021tensorflow}.
A model developed using the lighter TFLite framework is represented in a
different file format than that of the full TensorFlow model, called
\sw{FlatBuffers}\footnotemark. This portable, efficient, binary file format
offers a couple of advantages over using the \sw{ProtocolBuffers} model file
format, such as smaller file size through the reduced operations and code
footprint, as well as much faster inference by way of zero-copy deserialisation
for direct memory access without having to copy it into a separate part of
memory first for an additional parsing or unpacking step.
\footnotetext{\href{https://github.com/google/flatbuffers}{github.com/google/flatbuffers}}

For the most part, deep learning architectures are still designed and built
using the full TensorFlow framework, and only when the practitioner is
satisfied, is the model then \emph{converted} to a TFLite model, using the
helpful TFLite converter tool. The process of converting the original model to a
TFLite version is where many optimisations actually take place, with the
principle optimisation being operator fusion.

TensorFlow operations are themselves simple primitive operations which can be
combined together to form more complex operations. The primitive operations
appear as a single node in the computational graph that is constructed by
TensorFlow at runtime. Composite operations that are built from multiple
primitives, appear as separate nodes for each primitive operation. Fused
operations, on the other hand, act as a single operation that comprises of all
the computations that each primitive operation would normally have, as a single
node in the graph. By taking advantage of the underlying kernel implementations,
fused operations can maximize performance and reduce the code and memory
footprint, perfect for the resource constrained devices that it was designed
for, as well as for situations that demand low-latency inference, as in our
case. A useful by-product of fused operators is a high level interface that
helps define complex transformations such as quantization, that would otherwise
be cumbersome to map network wide.

\begin{figure}
    \centering
    \includegraphics[width=0.45\textwidth]{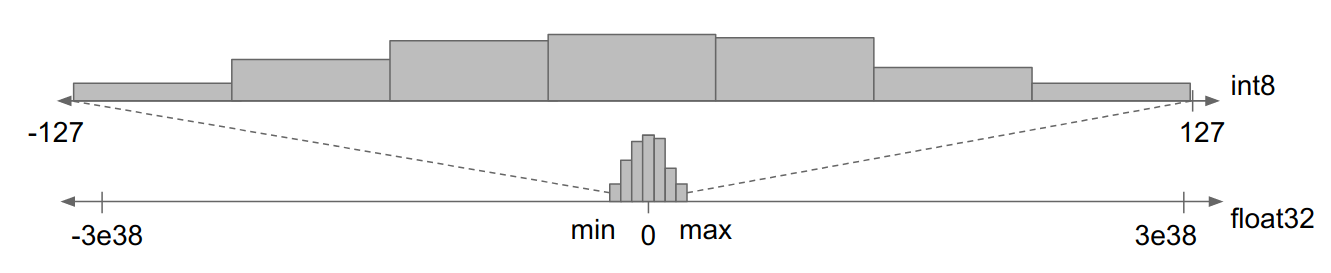}
    \caption[Quantization mapping of float representation to integer
    representation.]{Quantization mapping of float representation to integers
    representation. With the full range for 32-bit floating points extending
    from 3e\textsuperscript{-38} to 3e\textsuperscript{38}, there is often a
    remarkable amount of bits reserved for the precision, when in fact the
    majority of the numbers exists within a much narrow range on the number
    line. Integer numbers represented with 8-bits extend from -127 to 128 for
    signed values, and 0 to 255 for unsigned integers. With the appropriate
    mapping and scale factor, 32-bit numbers can be easily approximated as 8-bit
    integers, though 8-bit precision only has 255 information channels, this is
    a lossy conversion. Reproduced in full
    from~\citet{tensorflow2020quantization}}
    \label{figure:quantization}
\end{figure}

Under the TFLite framework, in conjunction with fused operations, quantization
is far easier to achieve, and actually appears as a simple flag at conversion
time when going from the original model to the TFLite version.
As described in Section~\ref{section:deep-compression}, quantization is the
procedure of mapping floating point values to a reduced integer representation
(see Figure~\ref{figure:quantization}), and in this TFLite setting, falls under
post-training quantization umbrella. Perhaps unique to TFLite, is that at
runtime, the model weights that are saved as integers, are scaled back to an
approximation of the original floating point values, to allow for computations
using floating point kernels to give better consistency with how inferences
would have resulted had the model had not been quantized in the first place. The
formulae for approximating floating point values from the saved integer weights
can seen here,
\begin{align}
  R = \left(Q - Z\right) \times S,
  \label{equation:quantization}
\end{align}
\noindent where the real value $R$ is approximated by a scale factor, $S$, that multiples the
difference between the $Q$-bits representation (which is commonly taken to be 8 for 8-bit integer
precision) and the zero-point value $Z$.

\begin{figure}
    \centering
    \hspace*{-1cm}
    \makebox[0pt]{\includegraphics[height=13.25cm]{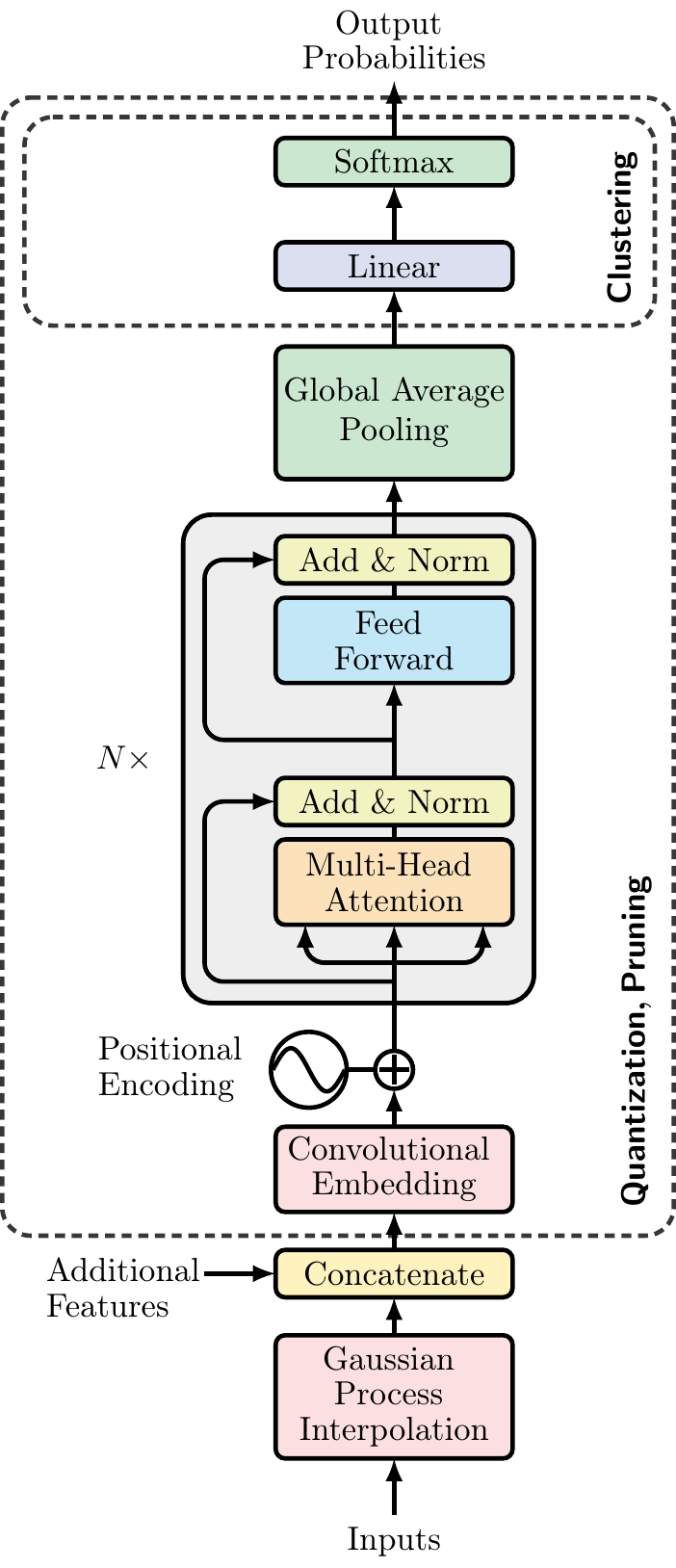}}%
    \caption[Locations within the time-series transformer (\texttt{t2})
    architecture, where deep model compression techniques have been
    applied.]{Locations within the time-series transformer (\texttt{t2})
    architecture, where deep model compression techniques have been applied. We
    investigate three forms of model compression, weight-pruning,
    weight-quantization and weight-clustering. Both weight-pruning and
    weight-quantization techniques are applied post-training, and are applied on
    all weights in the network. Weight-clustering on the other hand, is applied
    \emph{during} training, and only on the final dense layer to exploit
    parameter redundancy and to avoid the critical layers such as those in the
    attention block.} \label{figure:tinho}
\end{figure}

This section has described the compression schemes and optimisations applied to
our deep learning model to improve latency and reduce model size, yet with the
aim to preserve accuracy. Figure~\ref{figure:tinho} shows where each of these
techniques have been used in the time-series transformer architecture. Notably,
weight-clustering has not been used architecture-wide as it is not advisable to
cluster weights in critical layers early in the network\footnotemark.
\footnotetext{\href{https://www.tensorflow.org/model\_optimization/guide/clustering/clustering\_comprehensive\_guide}{tensorflow.org/model\_optimization/guide/clustering}}
However, as weight-pruning and weight-quantization are done post-training, we
are able to apply these techniques to the model as a whole.

\subsection{Hardware-Accelerated Distributed-Training}\label{deploy:section:hardware-distributed-training}

So far we have discussed optimisations that can be applied to the model in order
to save on disk space and improve runtime latency such that classification
scores can be given in real-time on the high-volume of incoming alerts. However,
of significant importance is the ability to quickly retrain and update the model
as new data becomes available. It is expected that FINK will have a window of 8
hours where new models will have the opportunity to be retrained before a new
round of LSST data ingestion and processing takes place~\citep{moller2021fink,
leoni2021fink}. Therefore, to have a model that can be retrained, and hence
improved with more refined data, within this window is highly desirable.

As with the case in FINK, which scales out computation across many CPU-only
machines in the Processing cluster to be able to churn through the large amount
of data quickly, we can take advantage of the same data parallelism principles
to also scale out computations for retraining models. By simply splitting the
dataset across multiple compute nodes, one can achieve a near linear-time speed
up. However, beyond scaling out to more CPU cores, we re-implement the
time-series transformer's training loop to scale out computations with multiple
hardware accelerators, in this case graphical processing units (GPUs), for
maximum speed up\footnotemark.
\footnotetext{We note that the codebase can easily be extended to run on even
faster tensor processing units (TPUs), but this was not taken further due to
lack of available resources.}
Compared to CPUs, GPUs do computations far more efficiently, saving time, energy
and costs in the long-term, but to ensure one takes full advantage of the
hardware accelerators capabilities, maximising the memory use at all times is
essential. If a GPU is underutilised, depending on the model and data input
size, severe training time degradation can be observed due to communication
overheads from host device (CPU) to the GPU accelerator. The main, and perhaps
most straightforward, way one can ensure maximum utilisation is to increase data
input \idx{batch size}. It is worth clarifying here that in this setting batch
size does not refer to the full dataset containing \emph{all} training samples,
rather it is a subset of training samples, equivalently called a
\idx{minibatch}~\citep{Goodfellow-et-al-2016}. We shall use the terms
interchangeably going forward.

Batch size in itself can have a major impact on model convergence, but it plays
a significant role when striving for optimal performance and computational
efficiency. Since TensorFlow uses 32-bit precision for floating point operations
on the GPU, model parameters take up 4 bytes each. Using this information, it is
possible to determine the largest practical batch size that can deliver maximum
GPU utilisation.

In addition to greater computational efficiency, larger batch sizes on the GPU
are also expected to yield slight classification performance gains.
If we consider that the standard error of the mean is estimated from $n$ samples
as $\sigma / \sqrt{n}$, with $\sigma$ as the standard deviation of the samples,
we can see that with larger $n$, one can obtain a more reasonable estimate for
gradients~\citep[p.~271]{Goodfellow-et-al-2016}.
While it would normally be the case that the non-linear scaling of gradient
estimates would invoke a trade-off between how many samples to use and compute
resources, such is the computational efficiency of GPUs that the limiting factor
becomes the amount of memory that can be used instead.
Therefore, by increasing the batch size to be as large as possible for a single
GPU, and then scaling this by the number of GPUs available, through an
\emph{all-reduce} operation when running our stochastic gradient optimisation,
we can realise the improved classification performance in addition to
computational improvements as well.

\section{Preliminary Results}\label{deploy:section:preliminary-results}

This section presents the preliminary results of applying the model
optimisations outlined in this article to the original time-series transformer
model~\citep{allam2021paying}. We first run local processing tests on real ZTF
alert packets to gauge the potential performance when deployed into the
production system of FINK (which is discussed in the next section).

\subsection{Model Retraining}

The original time-series transformer was trained using a single NVIDIA V100 GPU,
with 32GB of memory. Section~\ref{deploy:section:hardware-distributed-training}
explained how computational efficiency gains could be made by increasing batch
size.
As a first test, we extend the \sw{t2} codebase to allow for multiple GPU
training while ensuring the largest batch size possible is dispatched to each
GPU. Through the model profiling tool of \sw{model-profiler}\footnotemark~we
determine the best minibatch size to be $4096$ using the same GPU as before. Now
on the order of 100 times larger batch size compared to the original model
described in~\citet{allam2021paying}, we see far greater utilisation of the GPU.
\footnotetext{\href{https://pypi.org/project/model-profiler/}{pypi.org/project/model-profiler}}

Scaling out computations across many machines and increasing the global batch
size gave remarkable speed up, bringing training time down from approximately 8
minutes per epoch to 2 minutes per epoch, where epoch refers to one full forward
pass and one full backward pass of all the examples in the training set.
With an average convergence rate of 130 epochs, we bring overall model training
down from 17 hours to nearly 5 hours, now well within FINK's window of
opportunity for retraining new models nightly. Note that this is a full model
retrain, and simple fine-tuning can be done at a fraction of the time should
this be the preferred method of model updating. By leveraging data parallelism
in this way, where the data is distributed across multiple devices for training,
should even more hardware accelerators become available, we can bring this down
further still with a \emph{near}-linear reduction in training time.

\subsection{Local Processing Tests}\label{section:local-tests}

With the knowledge that we can retrain models quickly within the specified
window suitable for FINK nightly updates, we now move to our first application
of deep model compression.
We discussed in Section~\ref{section:deep-compression} that of the three
methods: weight-clustering, weight-pruning, and weight-quantization, only
clustering is applied during training. To enable this, we modify our original
model to allow for clustering of weights in the network.

Clustering, otherwise known as weight-sharing, can indeed be done throughout the
network. However, it is advisable to avoid highly specialised layers such as
attention blocks, and only focus clustering on the layers that are likely to
have high redundancies. For this reason, we only apply clustering to the final
dense layer, as shown in Figure~\ref{figure:clustering}.

\begin{figure*}
    \centering
    \includegraphics[width=0.95\textwidth]{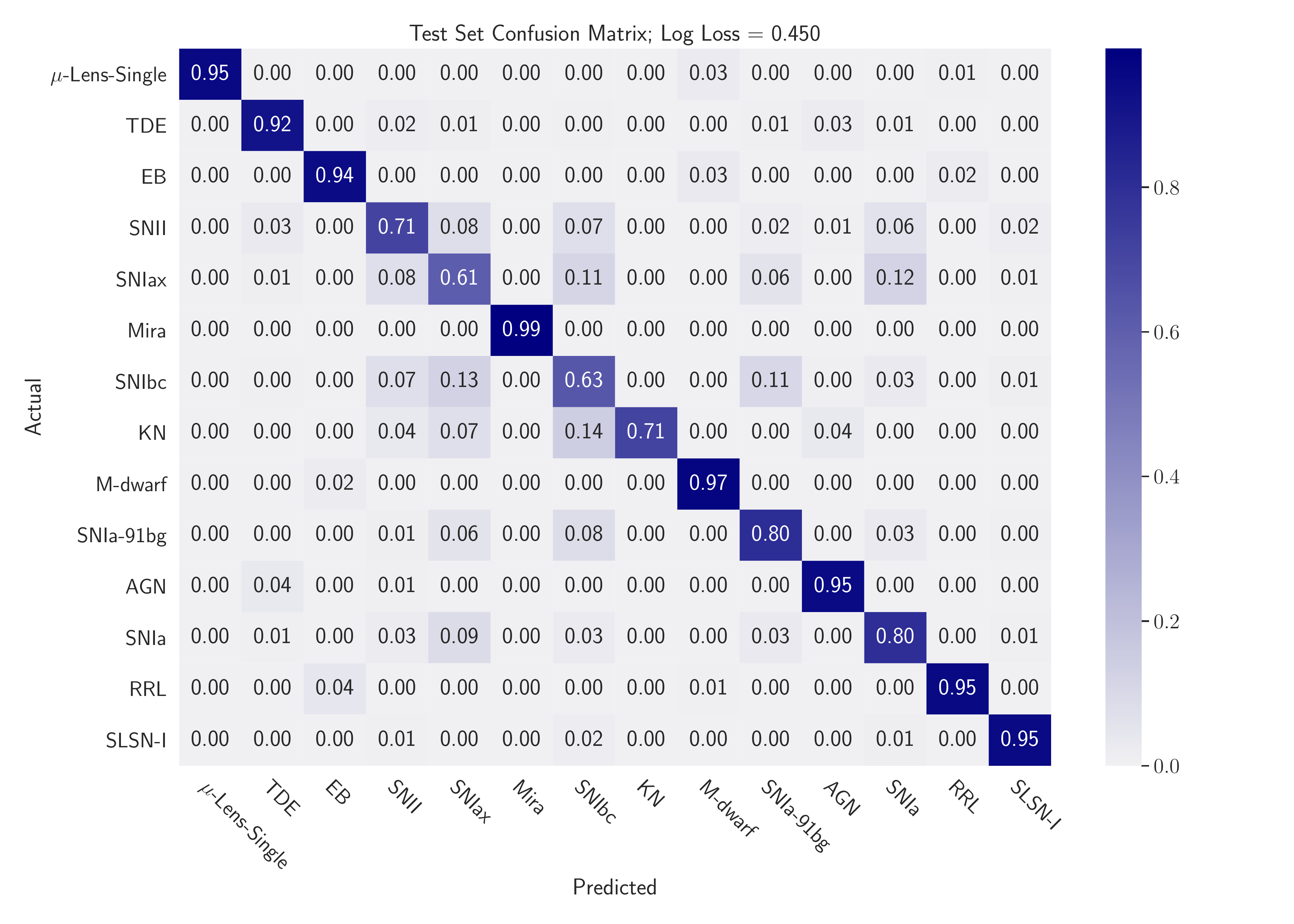}
    \caption[Confusion matrix resulting from application of a \emph{clustered}
    version of the time-series transformer~\citep{allam2021paying}, to the
    PLAsTiCC dataset in a representative setting with imbalanced
    classes.]{Confusion matrix resulting from application of a \emph{clustered}
    version of the time-series transformer~\citep{allam2021paying}, to the
    PLAsTiCC dataset~\citep{allam2018photometric} in a representative setting
    with imbalanced classes, achieving a logarithmic-loss of 0.450, using all 6
    passbands and additional information of redshift and redshift error}
    \label{figure:tinho-UGRIZY-wZ-confusion-matrix}
\end{figure*}

Firstly, we inspect the impact weight-clustering alone has on the model
performance compared to the original \sw{t2} architecture. Along with
application of model optimisations and compression, comes the expectation that
model performance could be adversely affected. While it may be the case that the
goal is to remove redundancies, by using these methods, there is inherent
information loss compared to the original model, which must be considered. Under
the same parameter settings and conditions as the original time-series
transformer model (see~\citet{allam2021paying}, Table 2), which used the six
passbands of $g, r, i, z, y$ plus two additional features photometric redshift
and photometric redshift error, we can see in
Figure~\ref{figure:tinho-UGRIZY-wZ-confusion-matrix} that by using clustering,
we not only preserve model accuracy, but this is improved slightly to a
logarithmic-loss of $0.450$ compared to $0.507$ previously. This actually takes
the performance beyond the previous state-of-the-art by~\citet{boone2019avocado}
of 0.468.
Furthermore, we maintain our ability to provide classifications at the level
required for cosmological analysis of dark energy with a
purity of 0.95 for SNIa and a core-collapse SNe (SNe Ib/c and SNe II) cross
contamination of $\sim 4.5\%$, comparable to results calculated for
DES~\citep{vincenzi2021dark} and Pan-STARRS~\citep{jones2018measuring} with
$\sim 8\%$ and $\sim 5\%$ respectively.
However, at this point, we should note that we are potentially seeing the
benefits from batch size enhancements and so slight improvements in
logarithmic-loss could perhaps be attributed to this, in addition to weight
clustering.

Nevertheless, redundancies have been exploited, allowing this model to achieve
good classification scores at a fraction of model size on disk. With
preservation of model performance confirmed, we continue to explore the impact
of applying other compression methods to \sw{t2} and inspect compression rate,
inference latency and model performance trade-offs.

To continue our preliminary analysis gearing for deployment, we simulate the
ingestion pipeline and use real ZTF alert packet data, such that it is akin to
what the model would be expected to handle in the production system of FINK.
To synthetically create a ZTF-\emph{like} dataset to retrain our model, we use
the same PLAsTiCC dataset~\citep{allam2018photometric} described in Section 5.1
of~\citet{allam2021paying}, using only the time-series information \ie~no
additional redshift features, and drop all passbands except for $g$ and $r$.
We then retrain to create a \emph{new} model that can handle ZTF data, and use
this as our baseline, which achieves a logarithmic-loss of $0.968$.
It should be noted that the synthetically generated ZTF dataset that is derived
from PLAsTiCC contains full light curves. This is generally not available in
real ZTF alert packet history data which only contain a candidate records for 30
days in the past. LSST on the other hand is set to provide up to 1 year of
historical data per alert.

In a comparative study, we look at four main aspects when judging machine
learning models for production performance: model size, model load time, model
inference time, and finally model performance in terms of logarithmic-loss
score. It is important to monitor any degradation in performance which would
indicate whether a particular technique, or combination of techniques are still
worth using. This is all shown in Table~\ref{table:compression-comparison},
which compares the baseline architecture of the original time-series transformer
described above with various compression methods and optimisations techniques
applied.

The first thing to note from Table~\ref{table:compression-comparison} is the
immediate space savings one can achieve through standard lossless compression
algorithms. The original model is able to be reduced down by $4.5\times$, which
is significant for space savings, but it is also clear that load and inference
latency are not affected. For the gains we are hoping for, deep compression
methods are required, and this will be the focus of the remainder of this
discussion.

The first method we apply is that of weight-clustering. We described in
Figure~\ref{figure:centroid-init} the typical ways to initialise the centroids
of the clusters. We opt for using linear initialisation and set the number of
clusters to be 16 for the reasons laid out in~\citet{han2015deep_compression},
which puts poor performance of the other initialisation procedures down to fewer
clusters containing large weight values.
The immediate effects of clustering show improvements in logarithmic-loss from
$0.968$ to $0.836$. We suspect this may be due to the reduction in the number of
parameters, helping to generalise the model as alluded to
in~\citet{lecun1989optimal}.
We also see a slight reduction in model size, but considering we now have shared
weights, the real model reductions are realised when we combine this with
Huffman encoding, which uses fewer bits for repeated values. Hence, we see a
better reduction in model size using clustering \emph{and} Huffman encoding
compared to just using Huffman encoding on the baseline model alone.
We would expect far greater compression rates should the technique be applied to
more layers than just the final dense layer which goes from 448 unique values to
16 values. Models which have many thousands of unique values would greatly
benefit from this procedure. In addition to space saving, we also see a small
reduction in load latency as well as inference latency.

We now move to the application of weight-pruning. Recall pruning removes
\say{unimportant} weights by setting them to zero, and specifically those
weights with low magnitudes. We apply this technique network wide so all layers'
weights are evaluated and trimmed down accordingly.
%
Through fine-tuning of the clustered network, we soon discovered that any
additional sparsity that was introduced negatively affected the classification
scores. The results can be seen in Table~\ref{table:compression-comparison},
where even pruning to a level of $1.1\%$ sparsity degraded performance. Further
empirical studies are necessary to determine what level of sparsity would
actually benefit this network.
Notwithstanding, we complete our analysis by including Huffman encoding to this
pruned network to witness any improvements in load time and inference latency.
Indeed there is an improvement on both of these metrics, but at the cost of
classification performance, we disregard using pruning any further.

We move towards a different approach for model optimisation with a change in
file format and framework as described in Section~\ref{section:file-formats}, as
well as application of weight-quantization. These results are denoted by the
dagger ($\dagger$) symbol in Table~\ref{table:compression-comparison}. These are
models that have originally been trained using the full TensorFlow framework but
converted to a more optimised, efficient file format of \sw{FlatBuffers}. The
conversion also involves operator fusion, to combine primitive computations that
appear as a single operation in the computational graph. With reduced code
footprint through operator fusion, and efficient binary representation of data,
we naturally see a large reduction in model size on disk. Around $10\times$
space savings can be achieved by simply converting the original model into a
TFLite model. In addition to the impressive space saving gains that are made,
with the model in this format we can take advantage of directly mapping the
model into memory for a reduction in load latency of more than $13,000\times$
speed up compared to the exact same clustered model and almost $15,000\times$
that of the original baseline.
As the model is loaded for each batch of data FINK processes\footnotemark, this
should lead to a fair increase in potential throughput of alerts.
\footnotetext{A batch, in this context, refers to the number of alerts bundled
together to then be distrubted across the cluster for processing. It is a
compromise between the number alerts simultaneously processed by each mapper in
the cluster, and the time delay between two alert stream polls: more alerts per
mapper leads to a more efficient computation by \emph{e.g.} minimising calls to
model loading, but it will increase the time delay between two batches, leading
to redistribution latencies. Therefore low latency model loading allows for more
freedom when choosing batch size in FINK, and how often to poll.}
While this would certainly help with throughput of alerts in the production
system, the other key metric for success is inference latency. That is, the time
go from alert packet ingestion to classification. It can be seen in
Table~\ref{table:compression-comparison} using the clustered model in the
\sw{FlatBuffer} format gives a speed up of around $5\times$ that of the same
clustered model, and around $7\times$ speed up compared to the original
baseline. Considering our model is expected to process millions of alerts per
night, having inference latency gains of this magnitude is undoubtedly positive.

Finally, we apply the third technique described in
Section~\ref{section:deep-compression} of quantization. To use this method, we
leverage the functionality that comes with TFLite's model conversion tool, that
allows for static quantization to 8-bit integers by examining the dynamic range
of the weights when saving model to disk, and then upscaling to floating point
approximation at inference time. By quantizing the weights of the clustered
model, and saving in \sw{FlatBuffer} format, we are able to shrink the model
even further to now $60$ kilobytes, an $18\times$ reduction when compared with
the original baseline model, and an incredible load time improvement of
$24,000\times$ speed up. Moreover, inference latency is reduced slightly
compared to the clustered model saved in \sw{FlatBuffer} format, for an overall
gain of nearly $8\times$ against the baseline. An important point is to mention
the preservation of model score with a logarithmic-loss of $0.834$.
Note the slight improvement in performance here compared to the clustered model
without quantization. We suspect this discrepancy between the other clustered
models to be due to the scaling approximation in
Equation~\ref{equation:quantization} and not due to the application of
quantization itself.

We have shown that application of compression techniques and use of appropriate
file formats, substantial space and memory savings, alert processing throughput,
and inference latency can be achieved.
However, we acknowledge local tests of the pipeline, while on real data, may not
be indicative of how well a model would perform in a real production systems,
under real-time constrains. Therefore, in the next section we put forward our
best performing model that uses a combination of clustering and quantization to
be deployed in a live setting on the production system of FINK for tests of
real-time classification.

  \begin{table*}
    \caption{Comparative performance between the original time-series transformer model, referred as
    the baseline, and the respective compressed versions using a combination of weight clustering, weight pruning and
    Huffman encoding. We present two sets of results in terms of models saved to disk in
    \sw{ProtocolBuffer} format and those saved in \sw{FlatBuffer} format, where the latter is
    denoted by a $\dagger$ symbol. Load latency refers to the time (in milliseconds) to simply read the model into
    memory, whereas inference latency (in seconds) tests the time to run predictions on a single ZTF alert
    packet. All tests were run on an Apple M1 Pro 32GB laptop.}
    \label{table:compression-comparison}
    \vskip 0in
    \begin{center}
      \setlength\tabcolsep{3pt}
      \begin{small}
        \begin{sc}
          \scalebox{0.95}{
            \hspace*{-2cm}
            \begin{tabular}{lcccc}
                \hline
                Compression Method                  & Model Size (kb)   & Load Latency ($\mathrm{s}^{-3}$) & Inference Latency ($\mathrm{s}$) & Loss \\

                \hline 
                Baseline                            & 1100              & 6324.145            & 0.333                   & 0.968 \\
                Baseline + Huffman                  & 244               & 6015.565            & 0.224                   & 0.968 \\
                \hline 
                Clustering                          & 892               & 5559.868            & 0.227                   & 0.836 \\
                Clustering + Pruning                & 688               & 5721.021            & 0.230                   & 1.017 \\
                Clustering + Huffman                & 240               & 4991.857            & 0.223                   & 0.836 \\
                Clustering + Pruning + Huffman      & 128               & 5251.288            & 0.228                   & 1.017 \\
                $\dagger$Clustering                 & 92                & 0.426               & 0.046                   & 0.836 \\
                $\dagger${\bf Clustering + Quantization}  & {\bf 60}    & {\bf 0.271}         & {\bf 0.043}             & {\bf 0.834} \\
                \hline 
            \end{tabular}}
        \end{sc}
      \end{small}
    \end{center}
    \vskip -0.1in
  \end{table*}

\section{Production Results}\label{deploy:section:production-results}

In the previous section we spoke of creating a new model that can classify ZTF
alert packets and hence be usable as a science module within FINK.
Comparing that to the original time-series transformer
of~\citet{allam2021paying} which worked with 6 photometric passband of PLAsTiCC:
$u, g, r, i, z, y$, as well as imputing additional features of photometric
redshift, we train a model derived from the time-series transformer architecture
that only takes in raw time-series from $g$ and $r$ bands, with no additional
features.
This is done to fit with the data that comes from ZTF alerts packets into FINK,
which do not contain photometric redshift information but only time-series
measurements for the two passbands of $g$ and $r$ filters.
It was then this $gr$-only model that was used as the baseline, shown in
Table~\ref{table:compression-comparison}.

\begin{figure*}
    \centering
    \includegraphics[width=0.95\textwidth]{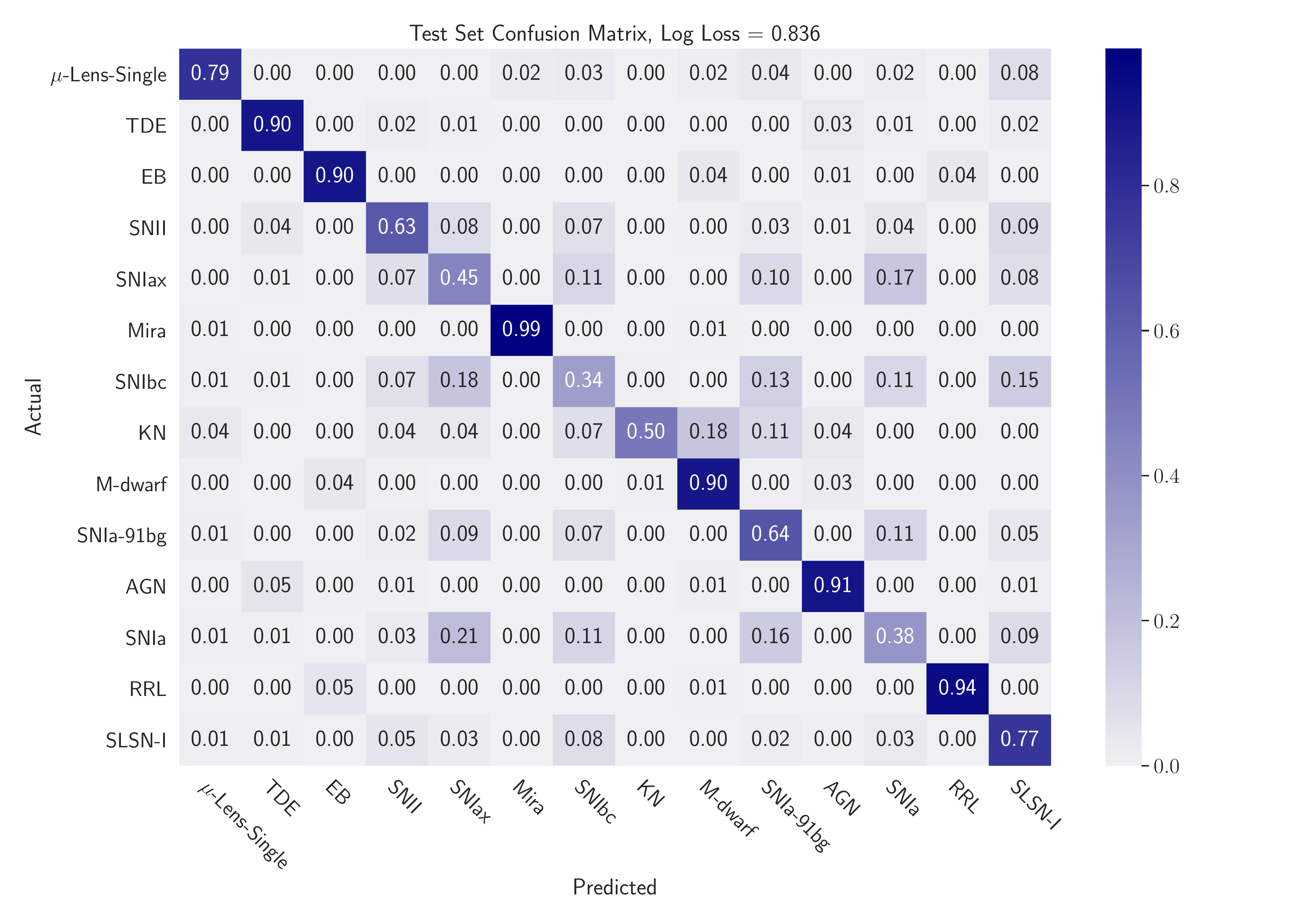}
    \caption[Confusion matrix resulting from application of a \emph{clustered} version of the
    time-series transformer~\citep{allam2021paying}, to the PLAsTiCC dataset in a representative
    setting with imbalanced classes, using only time-series information from $g$ and $r$ passband
    filters.]{Confusion matrix resulting from application of a \emph{clustered} version of the
    time-series transformer~\citep{allam2021paying}, to the PLAsTiCC dataset in a representative
    setting using full light curves, with imbalanced classes, and only time-series information from $g$ and $r$ passband
    filters. This model achieves a logarithmic-loss of $0.836$, using only the 2 passbands
  and no additional information.}
    \label{figure:tinho-GR-noZ-confusion-matrix}
\end{figure*}

Then, when using a combination of weight-clustering along with
weight-quantization saved in the more efficient format of \sw{FlatBuffers}, we
created the best performing model in terms of latency and space saving metrics
when tested locally on ZTF alert packet data.
The confusion matrix in Figure~\ref{figure:tinho-GR-noZ-confusion-matrix} shows
the performance of this quantized-clustered model trained on only 2 passbands.
We calculate a purity of $0.89$ for class SNIa, with a core-collapse SNe
cross-contamination of $8\%$, within the expected range described for
DES~\citep{vincenzi2021dark} and close to the $\sim 5\%$ described
in~\citet{jones2018measuring} for Pan-STARRS.

While the model is still able to make good classification scores across the
board, removing the other passbands of \emph{u, i, z} cause an increase in
cross-contamination by $\sim 4\%$.
This is interesting in its own right, where an avenue of research could lead to
investigate why training on only $g$ and $r$ passbands affect supernovae
classification in this way compared to when we can use all 6 passbands. This may
well be down to the lack of $i$-band specifically as this band along with
$r$-band is typically given preference in times of good seeing and at low
airmass~\citep{abell2009lsst}, but for our purposes, we just note these results
to keep in mind when assessing model validation.
It should be mentioned that we only train a model of this form to suit
deployment into FINK (that at the time of writing, can only ingest ZTF data),
for testing of our model as a real-time classifier. It is expected for LSST that
a model more akin to that showcased in
Figure~\ref{figure:tinho-UGRIZY-wZ-confusion-matrix} that uses all 6 passbands
as well as redshift information but trained on discretised alerts would be
deployed. Efforts are currently underway to develop and deploy such a model with
application to the \emph{Extended} LSST Astronomical Time-series Classification
Challenge (ELAsTiCC)~\citep{elasticc} and will feature in future
work\footnotemark.

\footnotetext{\href{https://portal.nersc.gov/cfs/lsst/DESC\_TD\_PUBLIC/ELASTICC/}{portal.nersc.gov/cfs/lsst/DESC\_TD\_PUBLIC/ELASTICC/}}

This section presents the results for deploying our \emph{quantized-clustered}
version of the time-series transformer model into the production system of FINK
tested on the real ZTF alert stream.
We compare the baseline model described in the previous section, that achieves
$0.968$ logarithmic-loss on ZTF-like data packets, with the compressed version,
that achieves better logarithmic-loss of $0.834$, in a now live production
setting, and observe alert throughput and latency improvements that have been
achieved when using the deep model compression techniques.

\subsection{Model Validation}

The first test for our deep learning model as a science module within FINK is to
validate the classification scores that we achieve. Not only is it important for
our model to operate in real-time under heavy work-load conditions, but it must
clearly continue to report correct classification results when deployed.
FINK validates models which target extra-galactic sources using the Transient
Named Server (TNS)~\citep{gal2021tns}, which is a transient alert system that
has spectroscopically confirmed objects in its database. By comparing
predictions to that of what TNS lists for a given object, we can get an
indication of how well a transient classifier is performing.

\begin{figure*}
    \includegraphics[height=15.5cm, width=0.85\textwidth, keepaspectratio]{{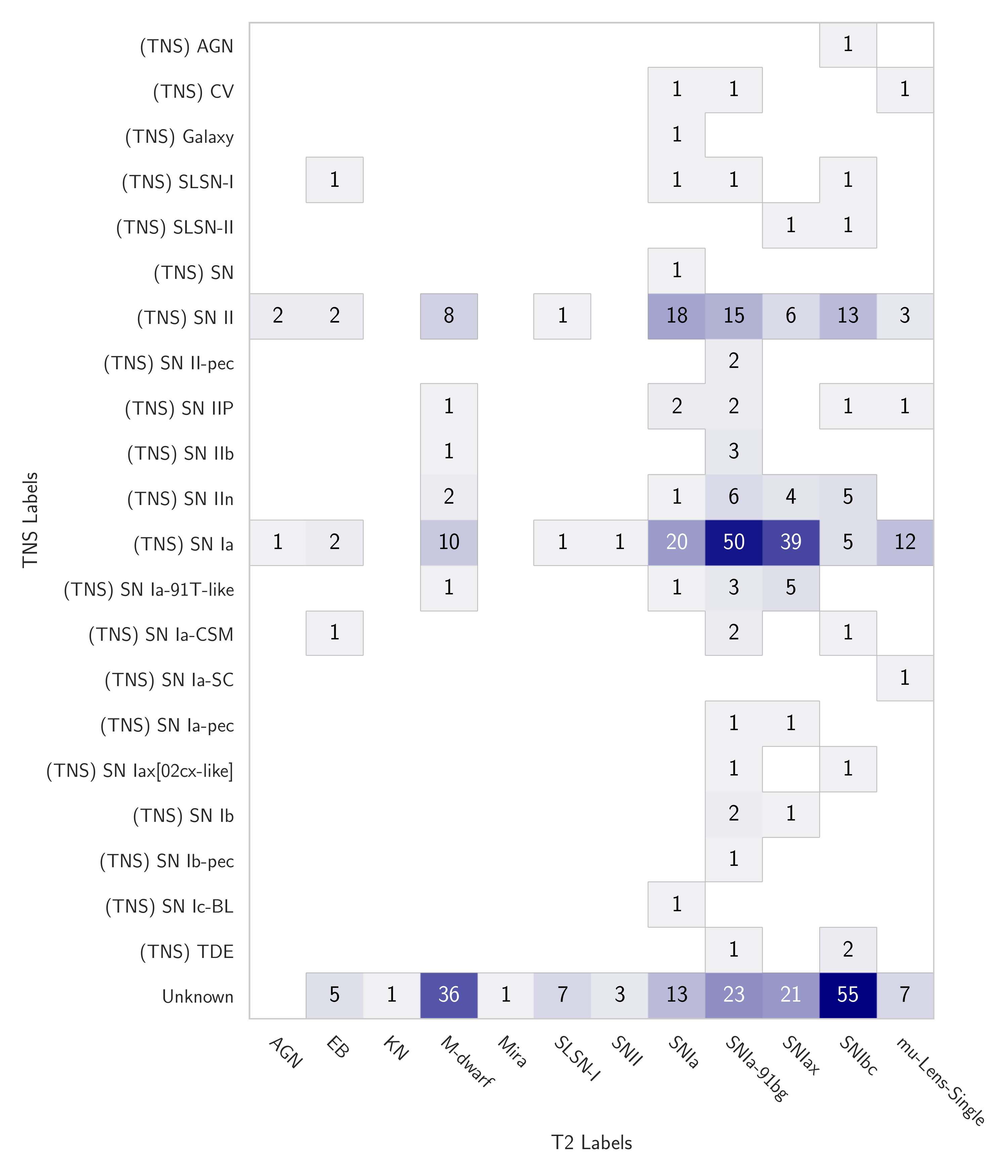}}
    \caption[.]{Comparison of spectroscopically confirmed labels in the
    Transient Named Server (TNS) database against the top-1 predictions for the
    compressed time-series transformer. The data used for this test comprised of
    one full year (2022) of real ZTF data with specific quality applied to the
    light curve history data requiring a minimum of 2 points and maximum of 90
    points on the light curve since the first alert emission date. The set of
    alerts are also reduced to filter out objects known to be a Solar System
    object from the MPC database or Galactic object when cross-matched against
    the SIMBAD database.} \label{figure:tns-validation}
\end{figure*}

\begin{figure*}
    \centering
    \includegraphics[width=0.8\textwidth]{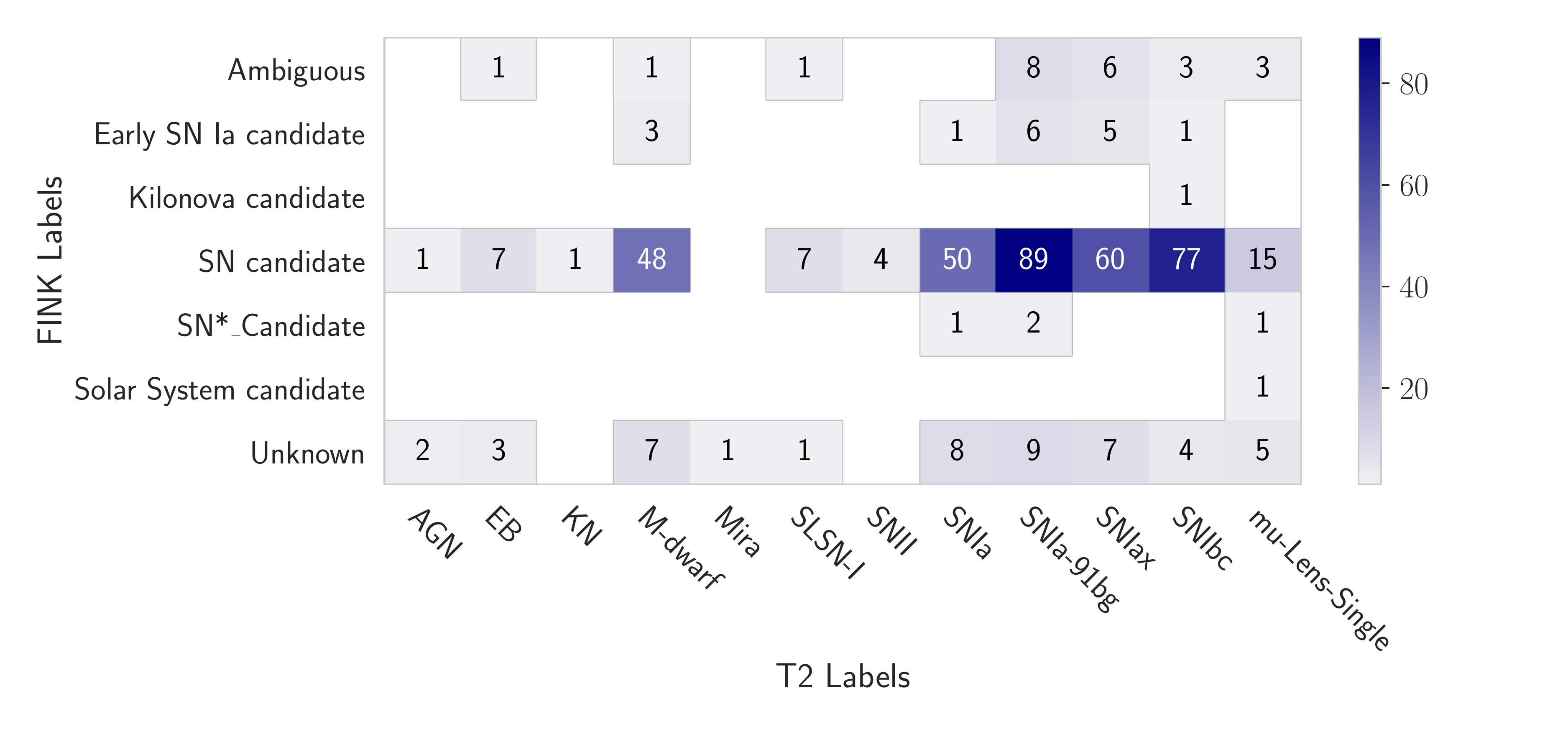}
    \caption[.]{Comparison of aggregated FINK classifiers' predictions against
    the top-1 predictions from the compression time-series transformer. One full
    year (2022) of real ZTF alert data is filtered according to the respective
    classifiers quality cuts as well as the criteria for a minimum of 2 points
    and maximum of 90 points on the light curve since the first alert emission
    date. We also ensure to disregard alerts that correspond to be a Solar
    System object from the MPC database or Galactic object when referenced
    against the SIMBAD database.} \label{figure:fink-validation}
\end{figure*}

Figure~\ref{figure:tns-validation} shows our models predictive performance on
one full year of real ZTF alerts against the spectroscopically confirmed objects
in the TNS database (extra-galactic objects). In addition to the quality cuts
common for all modules, we apply further criterion of at least 2 points and at
most 90 points on the light curve since the first alert emission date, as well
as the object to not be a Solar System object from the Minor Planet Center (MPC)
database or Galactic object according to the SIMBAD
database~\citep{wenger2000simbad}.
Recall in~\ref{section:local-tests} our baseline model has been trained using
full light curves. Since real ZTF alert packets contain only 30 days history,
only partial light curves are observable most of the time. This results in a
model being trained in non-representative setting where it expects full light
curves but is tested on discretised light curves. Yet, our model is able to
correctly identify the majority of SNIa objects, as well as other classes.
Though it should be noted there is greater misclassification of SNIa and
core-collapse SNe beyond the predicted cross-contamination of $\sim 8\%$
described above and this is likely due to the non-representative nature of the
training data the model has been built upon.
As such, we would not consider our model in this form to be suitable for a
fine-grain transient classifier, and ideally would need to be trained on the
discretised data of alert packets to be more representative, which is planned
for future work.
Considering these results, we can instead frame our model as to be a general
transient classifier that is able to identify SNe more broadly. Indeed, when we
evaluate the model against the ensemble of predictions from all other
classifiers in FINK we can see our model is able to correctly identify SN
candidates, shown in Figure~\ref{figure:fink-validation}.

Therefore, while there are some misclassification within SNe classes, the
compressed time-series transformer is able to successfully classify supernova
objects in general, and when compared to existing science modules in FINK,
correctly identifies supernova candidates. It is also able to go further, where
FINK science modules labels certain objects as \say{Unknown}, our model is able
to accurately suggest these as supernovae \emph{candidates} (see
Figure~\ref{figure:fink-validation}). A clear value-add for the brokering system
which can be used to update and enrich the database.

\subsection{Alert Throughput/Latency Performance}

With confirmation that the compressed model is operating correctly, we now come
to test the throughput and latency of the classifier, ultimately deciding the
usefulness of our model for real-time classifications.

We first look at how the original time-series transformer, which we refer to as
the baseline model in Table~\ref{table:compression-comparison}, fairs up against
other existing science modules in FINK. For this, we take one full nights worth
of ZTF alerts, amounting to approximately $200,000$ alert packets, and compare
the throughput and latency performances of all the science modules currently
implemented in \sw{fink-science}~\footnotemark as of \sw{v2.0.0}. For an overall
comparison, we show the \say{on-sky} throughput performance that passes all
alerts through all science modules, ignoring any pre-processing filters that
would normally be applied.
\footnotetext{\href{https://github.com/astrolabsoftware/fink-science}{github.com/astrolabsoftware/fink-science}}

\begin{figure*}
    \centering
    \includegraphics[width=0.85\textwidth]{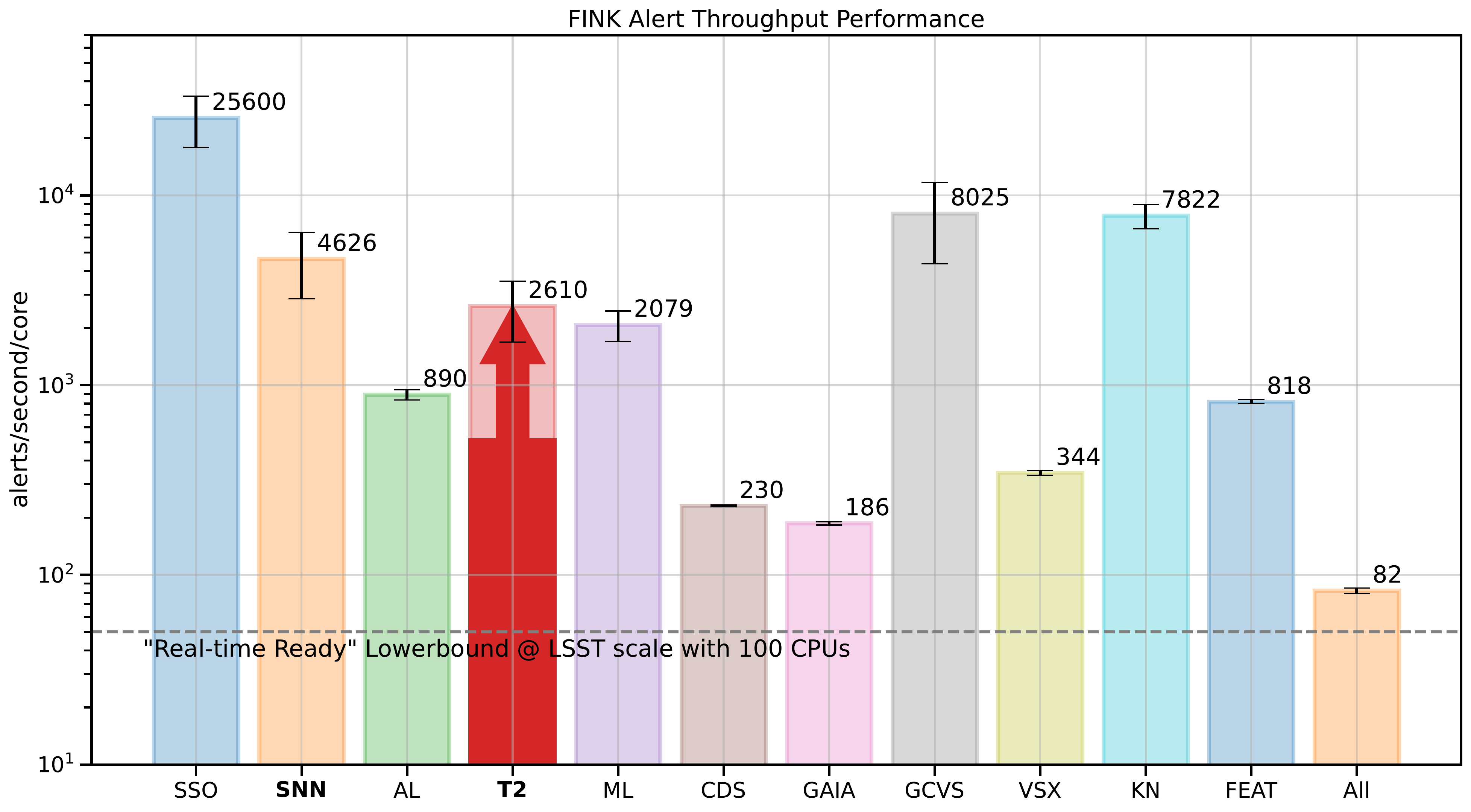}
    \caption[Improved alert throughput of the time-series transformer following
    application of deep compression techniques.]{ Alert throughput of FINK
    science modules, as of version \sw{fink-science-v2.0.0}, tested on one full
    nights worth of ZTF alerts (approximately $200,000$).
    As described in~\citet{moller2021fink}, we trace a horizontal line to
    indicate the threshold for a single science module within FINK to be
    considered \emph{real-time ready}, assuming 100 CPU cores. Under the data
    rates estimated for LSST, FINK will receive $10,000$ alerts every 37
    seconds, and such a threshold would allow for approximately a dozen science
    modules to provide classification scores serially. For full details of the
    inner workings of the other science modules shown here, the reader is
    advised to explore the \sw{fink-science} package.
    It should be noted that the results presented here are to be considered
    \say{on-sky} performance, where all alerts are processed blindly by all
    modules. This would not be the case typically, for example only a small
    fraction of alerts would be processed by the ML
    (microlensing)~\citep{godines2019machine} module since various inbuilt
    physics filters would determine if an alert is suitable for processing
    beforehand.
    Our time-series transformer model (T2), is originally able to achieve a
    throughput of $\sim 500$ alerts per second per core, but following
    application of deep compression techniques achieves an increased throughput
    of approximately $5\times$ of the number of alerts to now $\sim 2600$ alerts
    per second per core. In this same plot we have shown the baseline
    performance, and the new compressed version of the time-series transformer
    architecture using an arrow to indicate the increase in throughput. This
    version of the time-series transformer that uses weight-clustering and
    weight-quantization along with TFLite fused operations achieves performance
    far beyond the requirements for real-time classification of alerts at LSST
    scale. This has knock-on benefits for all other science modules within FINK,
    and encourages use of these methods for other science modules going forward.
    We highlight in bold along with T2 the only other deep learning model of SNN
    (\sw{SuperNNova})~\citep{moller2020supernnova}. The increase in throughput
    performance brings it within the same order-of-magnitude of alerts able to
    be processed, yet we are able to provide a probabilities score across all
    classes as opposed to a single binary classification. }
    \label{figure:improved-throughput}
\end{figure*}

Over an average of 20 processing runs, the mean alerts per second per core for
each science module is calculated. These results are most succinctly presented
in Figure~\ref{figure:improved-throughput}, where our model is the only deep
learning model of such complexity offering up a vector of probabilities for the
classification scores. Other science modules such as Solar-System Object (SSO)
and The Centre de {Donn\'es} astronomiques de Strasbourg (CDS) cross-matching
service are examples of table lookups whose performance is determined by the
execution of a query plan, and the only other deep learning model of
\sw{SuperNNova} (SNN)~\citep{moller2020supernnova} offers only a binary
classification for \gls{sn1}.

The baseline model, with no compression or optimisations made, is actually able
to sit amongst the other science modules and deliver real-time classifications.
While this seems to have already achieved our desired goal of deploying a
science module capable of real-time classification, it is important to consider
that the model that is deployed in FINK is not done so in isolation, but rather
all science modules within FINK will be operating in tandem.
Correspondingly, since the outgoing enriched alert packet is only sent after
\emph{all} science modules have finished processing the data packet, each
individual science module can have a large impact on the real-time science
capability of FINK as a whole. To not delay other modules that may be
time-critical such as for Gamma Ray Burst (GRB) detections, science modules need
to optimise throughput wherever possible, ultimately benefiting the entire
system.

Therefore, by going further and looking at how our best performing compressed
model manages to deal with the alert throughput in a live setting, we can see in
Figure~\ref{figure:improved-throughput} a sizeable improvement. While our local
processing tests gave up to $8\times$ speed up compared to the baseline model
for inference latency, in a real production environment, we achieve an
impressive $5\times$ in a live setting. It is suspected that a decline in speed
up compared to what was achieved in a local processing context can be attributed
to communication overheads in the cluster, where networking bandwidth becomes
the bottleneck in place of computations, as well as differences in hardware
capabilities.

This substantial throughput performance, thanks to low-latency inference via
deep compression techniques, greatly benefits the overall FINK system. As
science modules are run serially in FINK, our models ability to quickly complete
processing not only ensures there is no delay to other time-critical science
modules, but also permits more science modules to co-exist within the total
computational time budget afforded to FINK.
Finally, by improving latency in this way, we lay out a guide for other existing
deep learning models, and those under current development, for how to use model
optimisations for improved performance.

\section{Conclusions}\label{deploy:section:conclusions}

We have shown through deep model compression, complex models such as the
time-series transformer can be made super-lightweight for real-time inference.
The already efficient architecture benefits even further from weight-clustering
and weight-quantization to provide low-latency, high-throughput classification
scores, all the while preserving the accuracy of the results. Our study of
weight-pruning showed good reduction in model size but proved to be detrimental
to performance. Clearly even the low magnitude weights of the network carry
information critical for good classification.

We have shown through careful choice of file formats, major speed ups can be
achieved, which in turn dramatically improves a deep learning model's ability to
process inputs and operate in real-time, in a live production setting.
Our compressed version of the time-series transformer now resides in FINK as one
of the deployed science modules, operating at production scale providing nightly
classifications for the incoming ZTF alert stream since January 2023.
We have showcased our models suitability for providing robust classifications at
a fraction of the original model size and runtime. By scaling out computations,
we have brought retraining down to within the time frame required for nightly
updating on new alert data.

As described in Section~\ref{section:proxy}, the ZTF alerts stream, although
1/10\textsuperscript{th} of the expected LSST data rates, is a good precursor
for modelling the suitability of models and infrastructure to how well they will
handle future data streams. Consequently, we used FINK to emphasise our model's
ability to handle such large volumes of data and have presented results that
showcase its ability to cope with LSST scale, and beyond.

It is hoped that the work here, which introduces deep compression to the field
of real-time transient classifiers, will be harnessed to enable existing
architectures to be deployed as real-time classifiers easily into other
brokering systems, as well as to inspire those currently being developed that
real-time capability is within reach if techniques like those described here are
applied.


\section*{Acknowledgements}

The authors acknowledge the use of the UCL Myriad High Performance Computing Facility
(\texttt{Myriad@UCL}), and associated support services, in the completion of this work. The work was
also supported by the Science and Technology Facilities Council (STFC) Centre for Doctoral Training
in Data Intensive Science at UCL. This work made use of the FINK community broker resources. FINK is
supported by LSST-France and CNRS/IN2P3. TA would like to thank the FINK team for their helpful
discussions.

\section*{Data Availability}

All data referenced here can be found freely available online. For the original PLAsTiCC dataset,
the reader is advised to explore details laid out in \citet{allam2018photometric} and
\citet{team2019unblinded}. Code to reproduce the work found herein can be found at
\href{https://github.com/tallamjr/astronet}{github.com/tallamjr/astronet/tinho}.

\bibliographystyle{mnras}
\bibliography{deep-learning-deployment} 





\bsp	
\label{lastpage}
\end{document}